\documentclass[submit]{epsv8}
\usepackage{lineno}
\usepackage{graphicx}
\usepackage{amsmath}
\usepackage{float}
\usepackage{soul}
\usepackage{xcolor}
\usepackage{hyperref}
\title{Numerical study of multiple scattering intensities obtained from dust grains of different sizes and compositions in an impact-induced ejecta curtain}

\author{
Keerthana U, Manipal Centre for Natural Sciences, Manipal Academy of Higher Education, Manipal, India,
keerthana.mcnsmpl2024@learner.manipal.edu\\
P. Shalima, Manipal Centre for Natural Sciences, Manipal Academy of Higher Education, Manipal, India, shalima.p@manipal.edu\\
Koji Wada, Planetary Exploration Research Center, Chiba Institute of Technology, Tsudanuma, Narashino, Chiba-275-0016, Japan, wada@perc.it-chiba.ac.jp\\
Johannes Markkanen, Institute of Geophysics and Extraterrestrial Physics, TU Braunschweig, Mendelssohnstrasse 3, Braunschweig, 38106, Germany, j.markkanen@tu-braunschweig.de\\
Hiroshi Kimura, Planetary Exploration Research Center, Chiba Institute of Technology, Tsudanuma, Narashino, Chiba-275-0016, Japan, hiroshi\_kimura@perc.it-chiba.ac.jp 
}

\abstract{
Impact cratering of celestial bodies produces an ejecta plume containing underground material, which helps us understand the properties of subsurface grains and the surface evolution. In most cases, the ejecta plume forms an inverted conical structure so-called ejecta curtain.
In-situ measurements of light scattering of ejecta curtain, such as the images taken at the Hayabusa2's impact experiment help to constrain the size and composition of these less-processed grains using Monte-Carlo radiative transfer models. Here, we analyse the variation in the predicted scattered intensities of an ejecta curtain set up after that of Hayabusa2's case and test its sensitivity to different compositions, sizes, and phase function of grains comprising the ejecta curtain. Initially, the density/optical depth of the ejecta curtain is assumed to be constant throughout.  
The intensities are seen to deviate from single scattering expectations in some cases, which we attribute mostly to the effect of varying scattering angles and optical depths of the photons. Further, we show the importance of using the phase function reproducing actual grain materials in place of the often-used Henyey-Greenstein (HG) phase function in impact plume modelling. The results are compared with different values of optical depth in the ejecta cone and the study is also extended to incorporate a grain-size distribution. It is observed that there is an increase in the magnitude of the model median scattered intensity with the decrease of the power index of the grain-size distribution. Nevertheless, the overall variation in the profile between the different compositions remained the same as that of the monodisperse case with uniform density across the ejecta cone. Thus, we envisage this study to serve as a useful reference for investigating the subsurface grain properties of planetary bodies via impact plume observations, and to aid in improving available ejecta scattering models and also in the selection of prospective targets for future impact missions.
}


\begin{document}

\maketitle

\section{Introduction} \label{intro}

The study of the composition and size distribution of dust grains in planetary systems is one way to probe the formation and evolution of planetary bodies. Subsurface grains, which are mostly unaltered by the solar radiation, are a better source for studying the evolution of planetary bodies. Impacts are one of the ways to investigate subsurface grains. During the impact, subsurface grains are ejected during the excavation. The ejected grains collectively form an inverted conical structure, so-called ejecta curtain. The grain-size distribution and composition of the corresponding planetary body can be deduced by studying the scattered light profiles of the ejecta curtain. 

Spacecraft-induced impact experiments have been conducted in the Solar System such as asteroids, comets and the moon. As a part of the Deep Impact mission, the impactor collided with the surface of Comet 9P/Tempel1 and the event was observed by High Resolution Instruments (HRI) \citep{2005AHearn}. NASA's Lunar Crater Observation and Sensing Satellite (LCROSS) monitored the impact as well as the ejecta created by the upper stage of the launch vehicle near the south pole of the Moon \citep{2010Schultz}. The LADEE mission initially aimed to study the lunar thin atmosphere, finally completed its journey by creating an impact on the lunar surface \citep{2014Elphic}. Later in 2019, Hayabusa2 mission not only conducted an impact event using a small carry-on impactor on the surface of asteroid Ryugu, which is presumed to contain pristine material from the early Solar System. The impact event was captured by the deployable camera DCAM3 and evolution of the ejecta curtain over time is captured \citep{2020Arakawa}. The mission also brought back the samples in order to understand the formation and evolution of dust grains from the early Solar system \citep{2020Sugita}.  During the DART mission, the spacecraft hit one of the components of the asteroid binary system Dimorphos, the satellite of the asteroid Didymos \citep{2021Rivkin,2016Cheng}. 

Both Deep Impact and DART mission events have been modelled based on a 3D radiative transfer approach using publicly available software HYPERION \citep{2014Nagdimunov,2022Ludmilla}. A similar 3D radiative transfer model was developed by \citet{2015Shalima} considering cometary dust and ISM-type grains. This model is simple to handle and can be run on laptops or personal computers. Additionally, the model also uses the parallel rays of sunlight illuminating the ejecta curtain geometry. The direction of illumination irrespective of isotropic behaviour of the source makes the parameter study more powerful, along with the variation of other parameters in a reasonable amount of time. 

Constructing a 2D intensity map of multiple scattered intensity of an ejecta curtain as an `ejecta curtain scattered intensity map' provides a unique opportunity to predict the scattered intensity for a given set of sizes and compositions of the dust grains corresponding to the Solar System bodies. In this paper, we have taken a set of grain sizes and compositions commonly observed in the Solar System. Section \ref{model} discusses the calculation of cross-sections and scattering matrix elements in different size regimes, considering a 3D geometry of an ideal ejecta curtain based on Hayabusa2's impact event, along with the complete photon-dust interaction system. The section also includes the radiative transfer technique used to calculate the scattered intensity. This is followed by the obtained results and discussion in consecutive sections using the mentioned method.


\section{Ejecta curtain model}\label{model}

The general scattering model can be divided into three parts: the Sun-ejecta geometry which is fixed for the given physical system, the dust grain model where the dust parameters are constrained and the Monte-Carlo radiative transfer method which is used to obtain the scattered intensity map.


\subsection{3D Geometry}\label{sgeo}

\begin{figure*}[h]
\centering
\includegraphics[width=0.95\textwidth]{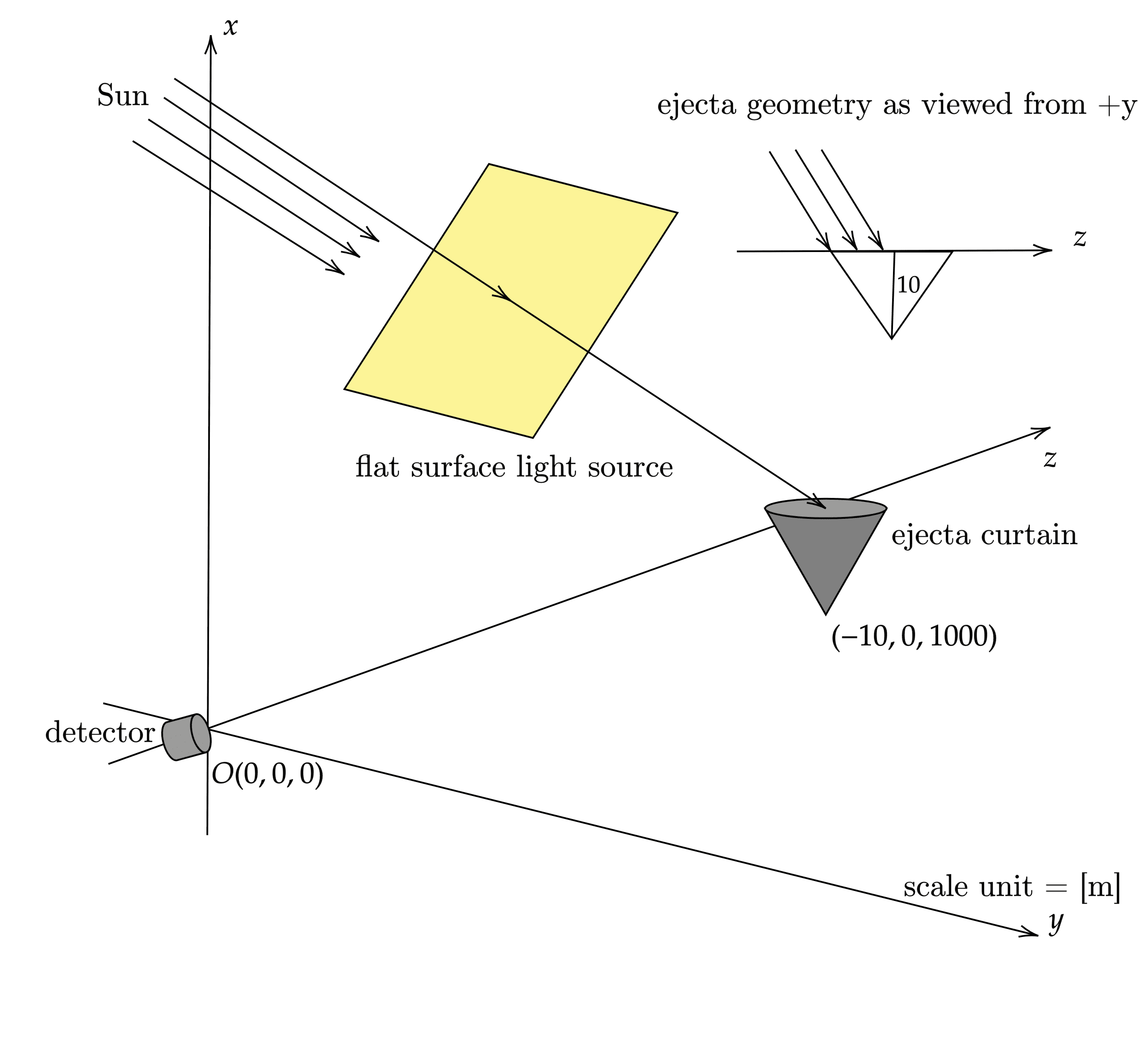}
\caption{Relative 3D distribution of the ejecta curtain, the direction of Sun rays and the detector. The parallel rays of the Sun is simulated by the flat surface light source \cite[see][]{2015Shalima}.}
\label{geometry}
\end{figure*}

In this study, we set up an ejecta curtain based on the Hayabusa2's impact event case so that we can check our calculation in comparison with the images of the impact event taken by the miniature high resolution camera DCAM3 \citep{2017ogawa}. Our model predicts the scattered intensities of the ejecta curtain in the visible range (6000 \AA), which corresponds to DCAM3 specifications, that would also be applicable for future planetary exploration missions with similar objectives. The assumed configuration of the ejecta curtain, the direction of sunlight and the detector is shown in Figure \ref{geometry}. In this study, we consider an inverted cone as the ejecta curtain structure for simplicity, although the ejecta curtain actually forms a cone truncated at the surface of impacted target. Here, the sunlight is incident on the ejecta curtain as parallel rays, at 45$^\circ$. This results in the phase angle range of 40$^\circ$ to 50$^\circ$ in the multiple scattering modelling. This is approximately equal to the phase angle (50$^\circ$ - 53$^\circ$) reported, whereas the incident angle is slightly higher than the reported range (36$^\circ$ - 41$^\circ$) mentioned in \citet{2026SKikuchi}. The incident photon flux corresponding to the wavelength $\lambda$ = 6000 \AA \ at 1 AU i.e., 1.3 $\times 10^{21}$ photons s$^{-1}$ m$^{-2}$ \AA$^{-1}$ is used in the calculation. It is assumed that there is no inclination of the ejecta curtain normal to the surface of the planetary body considered. The coordinates corresponding to the virtual apex of the curtain is (-10, 0, 1000) and the detector is placed at the origin (0, 0, 0). This results in the distance of the detector to be of 1000 m and the ejecta curtain is observed perpendicular to the ejecta cone axis which is similar to that of DCAM3 observations \citep[see][for the detailed description of Monte-Carlo method used]{2015Shalima}. It is essential to note that the scattered intensity can be easily scaled to the distance between the impact event and the detector and can be used for arbitrary observational distances. 

Assuming the typical ejecta curtain formed after a few seconds of impact on the surface of the planetary bodies \citep{2011Holsapple}, we have considered the geometry of the ejecta curtain to be a hollow cone of height and radius, 10 m each. The ejecta cone is made up of small cubes, of side 0.1 m. Each cell is defined by optical depth $\tau = \rho C_{ext} dx$, where $\rho$ is the number density of the grains in each cube, $C_{ext}$ is the extinction cross-section and $dx$ is the length of each cube. The number density of the grains in each cube is kept uniform throughout the ejecta cone. We have considered both monodisperse dust grains as well as the dust grains following power-law size distribution in the ejecta cone. We have also varied number density following a power-law to study overall parameter dependencies on the scattered intensity. 


\subsection{Dust grain model}

The 2D intensity maps are built on the foundation of a large combination of dust parameters. We have mainly concentrated on the parameters such as size and composition. The range of dust parameter values is chosen based on careful examinations of the literature on planetary bodies. 

\textbf{Size and structure:} The study of dust particles of comet Halley using the Giotto and Vega spacecraft suggested that the dust grain size ranges from 0.1 $\mu$m to 1 cm \citep{1986Tamas,1986McDonnell}. \citet{2014Nagdimunov} estimated the best-fit parameters for dust ejecta particles of Comet 9P is between 0.1 $\mu m$ to 100 $\mu m$. The typical size range taken to model the dust environment of comet 67P based on the observations of the Rosetta spacecraft and OSIRIS camera is 10 $\mu$m to 1000 $\mu$m \citep{2018Moreno,2018JohannesM,2019Guttler}. It has become a known fact from the recent observations of the asteroid Ryugu through the Hayabusa2 mission that the typical size of the particles can be of millimetres also \citep{2021KojiWada}. \citet{2023graykowski} reported the maximum brightness, followed by the reddening and fading of the ejecta plume brightness over a course of time, also estimated the lower end of the grain size to be 0.1 $\mu m$. The early observations of the DART impact ejecta plume using VLT/MUSE suggests that the initial ejecta may contain dust grains of size 1 $\mu m$ to 10 $\mu m$ \citep{2023opitom}. The HST observations suggests that the tail of the impact follows the broken power-law size distribution \citep{2023JYLi}. Using the radiative transfer modelling, the dust in the extended tail of an ejecta cloud at the asteroid after DART impact suggested that the best fit dust grain size ranges between 10 $\mu m$ to centimetre \citep{2025Ludmilla}. Considering these estimates, the size (radius) of the dust grain in this study is varied from 0.1 $\mu$m to 1000 $\mu$m.

The optical properties of the dust grains can be studied using Rayleigh theory, Mie theory and geometric optics. Here, we assume spherical dust particles simply to avoid the complexity that adds up to the calculations. The exact analytical solution of Maxwell's equations \citep{1908Mie} as given by Mie can be used to calculate the light scattering properties, such as cross-sections, scattering matrix elements for the considered homogeneous spheres. As Mie theory allows us to take particle size which is either comparable or larger with respect to incident wavelength also, where Rayleigh theory fails. Here we have used a Python package called {\fontfamily{qcr}\selectfont miepython} \citep{2025prahl}, which provides an efficient way to obtain cross-sections and scattering matrix elements for a wide set of dust parameters \citep{2023Larissa,2025Chen}.

\textbf{Composition: }The material is generally characterised using a complex refractive index, $m = n + ik$, with $n$ and $k$ being its real and imaginary parts, respectively. We have chosen compositions based on the literature survey of the commonly observed materials in the Solar System, with a particular interest on small planetary bodies as listed in Table \ref{comp}. Even though the most abundant elements in small planetary bodies are silicates and carbonaceous materials, a variety of other mixtures are also observed. 

The refractive indices of solid amorphous carbon were taken from the standard calculations using thin film across the wavelength 1176 \AA \ to 6666 \AA\citep{1984Duley}. The optical properties of graphite and astronomical silicates are well studied based on the available laboratory data, interstellar extinction observations along with Kramers-Kronig relations \citep{1984BTDraine}. Using the reflection and transmission functions, $n$ and $k$ values are determined for a given spectral range from 300 \AA \ to 1000 $\mu m$ using a metal thin films of graphite and astronomical silicate. Similar method is used to calculate the refractive index of iron for an energy range 0.64 eV (1.64 $\mu m$) to 6.6 eV (0.188 $\mu m)$
, which is commonly seen in meteorites. It is observed that the major constituents of Halley's dust are magnesium silicates, iron sulfides and other organic refractory \citep{1988Jessberger,1999Jessbrger}. By taking one-third of organic material and two-thirds of amorphous carbon containing silicates, iron etc. the average refractive index of cometary dust is calculated at two different wavelengths $\lambda$ = 4500 \AA, 6000 \AA. This composition mixture is also equivalent not only to the composition of Halley's dust, but also Interplanetary Dust Particles (IDPs) \citep{2003HKimura}. The refractive index of model particles which are equivalent to that of impact generated lunar glass is also considered at 6000 \AA \ \citep{2010Goguen}. The refractive index of Magnesium-rich Olivine reported at 4500 \AA \ and 6000 \AA \ is taken from \citet{2006HKimura}. Using inhomogeneous deposition of organic refractory residue, along with Kramers-Kronig relations, the optical constants are calculated for the wavelength range 0.115 $\mu m$ to 1.7 $\mu m$ \citep{1993Jenniskens}. \cite{1984BNKhare} conducted several experiments to replicate the Titanian atmosphere through simulations, obtained optical constant values for a range of wavelengths from 0.0250 $\mu m$ to 1000 $\mu m$ using thin film of Titan tholin.

From the above literature study, the chosen compositions almost cover a wide range of dust grains observed in the small planetary bodies, with a wide range of absorptive powers. The complex refractive indices listed in Table \ref{comp} are reported over a range of wavelengths in most cases, that
encompasses 6000 \AA and was obtained by interpolation within the wavelength range, whereas the remaining values were available at 6000 \AA. The weakly-absorbing materials like lunar glass and Mg-rich olivine have an imaginary part of complex refractive index $k \sim 0$. The materials that have moderately absorbing properties at 6000 \AA \ like astronomical silicates, organic refractories and tholin have $k$ ranging from $\sim$ 0.02 - 0.1. The strongly-absorbing materials like amorphous carbon, graphite, iron and cometary dust have $k$ in the range $\sim 0.5 - 3$. 

\begin{table}[h!]
\centering
\caption{Complex refractive index ($m = n + ik$) for different types of materials at $\lambda$ = 6000 \AA.\label{comp}}
\begin{tabular}{c c c} 
\hline
\hline
Composition & $m = n + ik$ & Reference \\
\hline
Amorphous carbon & 2.43 + i 0.59 & \cite{1984Duley}\\
Astronomical silicate & 1.68 + i 0.03 & \cite{1984BTDraine}\\
Cometary dust & 1.98 + i 0.48 & \cite{2003HKimura} \\
Graphite &  2.63 + i 1.13 & \cite{1984BTDraine}\\
Iron & 2.88 + i 3.05 & \cite{1974Johnson} \\
Lunar glass & 1.65 + i 0.003 & \cite{2010Goguen}\\
Magnesium rich Olivine & 1.68 + i 1.15 $\times 10^{-5}$ & \cite{2006HKimura} \\
Organic refractory & 1.57 + i 0.1 & \cite{1993Jenniskens}\\
Tholin & 1.70 + i 0.023 & \cite{1984BNKhare}\\
\hline
\end{tabular}
\end{table}


\subsection{Scattering Radiative Transfer Method}\label{MCRTmethod}

In a realistic scenario, the photons undergo multiple scattering and the Monte Carlo method is mostly used in such cases. The method was initially used to explain the scattering in reflection nebulae \citep{1977Witt}. Later, the method has been used in multiple astrophysical scenarios \citep{1996wood,2001Gordon,2002wood}. The computational method used here is similar to that of \cite{2004Shalima,2015Shalima}.  

\begin{figure*}[h]
\centering
\includegraphics[width=1.0\textwidth]{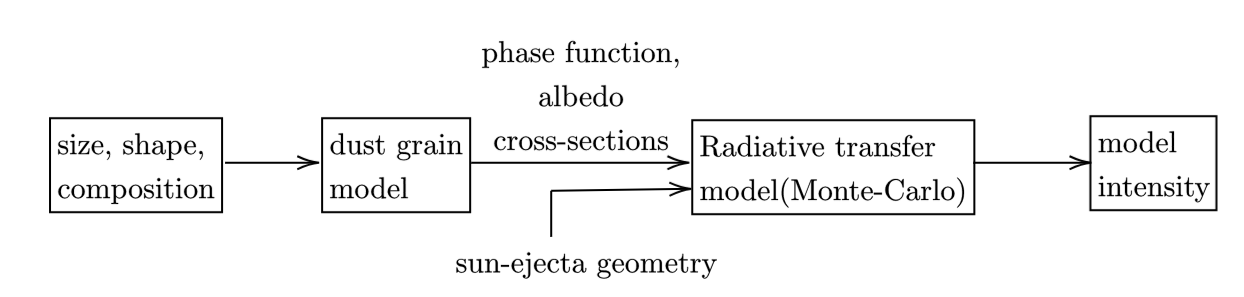}
\caption{General framework of scattering radiative transfer model.}
\label{RTmodel}
\end{figure*}

The method is based on the principle of tracking photon trajectory from the first scattering till it leaves the medium or its intensity drops significantly. The photon trajectory mainly depends on the scattering angle as well as the path length between consecutive scatterings. Both these parameters are dependent on the scattering phase function based on which they are uniquely determined. Using the concept of weighted photons, the photon's weight is reduced by the factor of single scattering albedo after each interaction. A part of energy of the scattered weighted photon is redirected to the detector. This fractional photon is further reduced by $e^{-\tau}$, where $\tau$ is the optical depth, when it interacts with the material along the path of the detector. Here, we have considered $10^7$ weighted photons for the calculation as the calculated model intensity converges to the solution after ten million iterations within the available computational limitation. During the calculation period, the required computational time is two and a half hours to generate an intensity map for a fixed set of dust parameters, 352.5 successful hours in total. 

The scattering phase function, which depends on the size, composition and shape of the dust grains, along with the wavelength of the incident light is used in the model. Assuming the sunlight is unpolarized, the phase function $\Phi(\theta)$ can be taken as \citep{1983Bohren},

\begin{equation}\label{phasefn}
    \Phi(\theta) = \frac{1}{C_{sca}} \frac{dC_{sca}}{d\Omega} = \frac{1}{C_{sca}} \frac{S_{11} (\theta)}{\text{k}^2}
\end{equation}

where $dC_{sca}/d\Omega$ is the differential scattering cross-section, $C_{sca}$ is the scattering cross-section, $\theta$ is the scattering angle, $S_{11}$ is the scattering matrix element and wave number k$ = 2 \pi / \lambda$, $\lambda$ is the wavelength of the incident light (6000 \AA). 

The scattered intensity ($I_s$), which depends on the single scattering albedo ($a$), phase function ($\Phi(\theta)$) and the intensity of the incident light ($I_i$) is given as

\begin{equation}
    I_s \propto a \ \Phi(\theta) \ I_i 
\end{equation}

Single scattering albedo gives the amount of scattering at a given wavelength and is calculated by taking the ratio of scattering cross-section to extinction cross-section obtained using Mie theory. 

The general steps taken to model ejecta curtain is shown in the Figure \ref{RTmodel}. Initially, the size, shape and composition of the dust grain at a particular wavelength are fixed. These fixed dust parameters are given as an input to the {\fontfamily{qcr}\selectfont miepython} package. The scattering cross-sections, asymmetric parameter and scattering matrix elements are obtained using this software. The geometry of the ejecta curtain, having a constant number density of the grains, is built using the Python programming language. The phase function values are calculated using the equation (\ref{phasefn}). The calculated phase function values, albedo, cross-sections of the dust grain, along with the ejecta curtain geometry is given as an input to the Monte-Carlo model developed by \citet{2015Shalima}. In the model, a fractional amount of scattered energy is directed towards the detector (visible camera) after every scattering, composed of 2000 $\times$ 2000 pixel grid with a resolution of 0.01$^\circ$/pixel, at a distance of 1000 m from the ejecta curtain and a 2D model intensity image is obtained [ref. Figure \ref{scatmap}]. The phase function, the scattered intensity map and the intensity profile obtained at each combination of dust grain parameter is shown as scattering tadiative transfer grid (SRTG) in the Appendices \ref{SRTGconst}, \ref{SRTGvar}.

The procedure is repeated, incorporating size-distribution and number density distribution in the ejecta curtain. Here, we have considered the power law distribution for both dust grain size and number density calculations. The dust grain size is distributed, with power-law indices of -2, -3 and -4. The limits of the dust grain size are taken as 0.1 $\mu$m and 1000 $\mu$m. The number density decreases from the base to the top of the ejecta cone, with the power-law index of -1.82. The optical depth values vary from 66.0884 $\pm$ 7.1252 to 0.0152 $\pm$ 0.0016 for the power index -2; 61.0209 $\pm$ 7.2364 to 0.0140 $\pm$ 0.0016 for the power index -3; 60.5467 $\pm$ 7.2474 to 0.0139 $\pm$ 0.0016 for the power index -4, at a fixed size distribution and averaged over all compositions.  


\section{Results}\label{result}

\subsection{Scattering phase function}

\subsubsection{Dust grain size dependence\\}

\begin{figure}[h]
\centering
\includegraphics[width=0.7\textwidth]{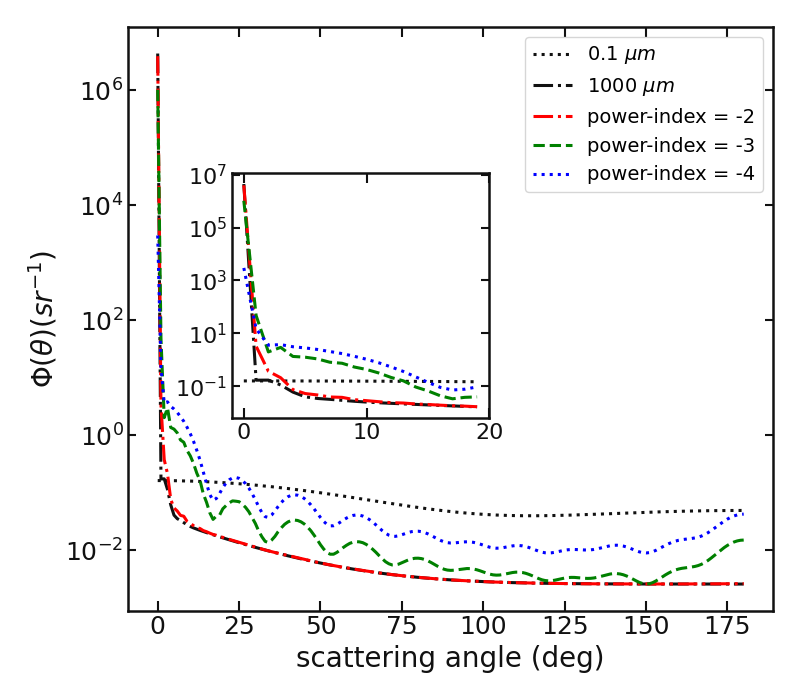}
\caption{Variations in phase function values with scattering angle for astronomical silicate grains of sizes, 0.1 $\mu$m, 1000 $\mu$m and for size-distribution with power indices -2, -3 and -4.}
\label{pfsize}
\end{figure}

Here, we show the calculated phase function values as a function of scattering angle for astronomical silicate grains of different sizes (see Figure \ref{pfsize}). The figure also shows the comparison between the phase function for individual dust grain size to that of the grain size distributed case with different power-indices values.

The phase functions are nearly constant at sub-micron size, implying grains are mostly isotropic as expected. The complexity in the profile of phase functions increased with the dust grain size, along with the series of maxima and minima of the phase curve across the Mie resonance range. It is also seen that the maximum values of the phase functions are restricted to smaller scattering angles at comparatively larger dust grain size, indicating strong forward scattering. This result matches the earlier results, where large grains showed strong forward scattering \citep{2000Munoz}. 
The obtained phase curve profile is similar to that of axially symmetric aggregates \citep{2000Petrova} and aggregates that are formed either under Ballistic Cluster-Cluster Aggregation or Ballistic Particle-Cluster Aggregation processes \citep{2003HKimura}, when dust grain size is comparable to the wavelength. We have also found similar dependence of phase function values on the dust grain size. On the other hand, the phase function corresponding to the dust grain-size distribution with power-law indices -2, -3, and -4 is shifted toward the phase function profile of the large dust grain size. Higher the power-law index, profile of phase function is similar to the larger dust grains as shown in the Figure \ref{pfsize}. 


\subsubsection{Dust grain composition dependence\\}

\begin{figure}[h]
\centering
\includegraphics[width=0.7\textwidth]{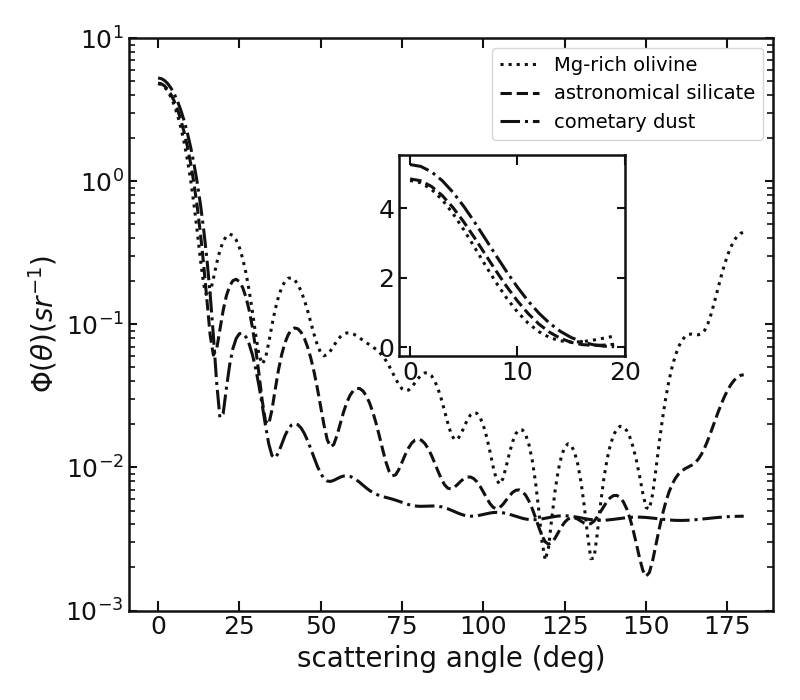}
\caption{Variation of phase function values with scattering angle for 1 $\mu m$ size Mg-rich olivine, astronomical silicate and cometary dust.}
\label{pfcomp}
\end{figure}

As shown in Figure \ref{pfcomp}, the obtained phase curves are dependent on the refractive indices of the compositions listed in the Table \ref{comp}, along with the dust grain size. The low absorbing property (\textbf{$k$} $\sim$ 0) of lunar glass and magnesium-rich olivine has shown ripple structure and interference pattern in the phase function profile, which is absent in the absorbing materials like iron, graphite etc. This is mainly due to its low absorptive power, which allows light to penetrate through the material and the interaction results in an interference pattern. The interference pattern and the magnitude have been reduced with the increase of absorptive power in the obtained phase curve. Here, three different types of materials based on their absorptive power, namely astronomical silicate, cometary dust and Mg-rich olivine are considered. The decrease in magnitude as well as the difference in the phase function profile, with the increase of absorptive power, is shown in the Figure \ref{pfcomp}. The result of weakly-absorbing materials are similar to that of the light scattering study done on non-absorbing particles using the Ramsauer approach \citep{2012Karim}. In general, the results show a significant difference in the nature of the phase function profile depending on their absorptive power of the material.

\subsubsection{Comparison between Henyey-Greenstein (HG) and the calculated phase function\\}

The Henyey-Greenstein phase function is mostly used model parameter to compare with the observations in multiple astrophysical scenarios \citep{1941HG,1977Witt,2003Hegmann}. In the current study, we have used the scattering phase function obtained from {\fontfamily{qcr}\selectfont
miepython} as explained in Section \ref{MCRTmethod}.
The HG phase function is given by the following equation (3) with an asymmetry parameter g, which is obtained from the calculation of {\fontfamily{qcr}\selectfont
miepython} [see Table \ref{gparameter} and \ref{gparameter_sizedistr}].

\begin{equation}\label{HG}
    \Phi(\theta) = \mathbf{\frac{1}{4\pi}}\frac{1 - g^2}{(1 - 2 g \cos{\theta} + g^2)^{3/2}}
\end{equation}

\begin{figure}[h]
\centering
\includegraphics[width=0.7\textwidth]{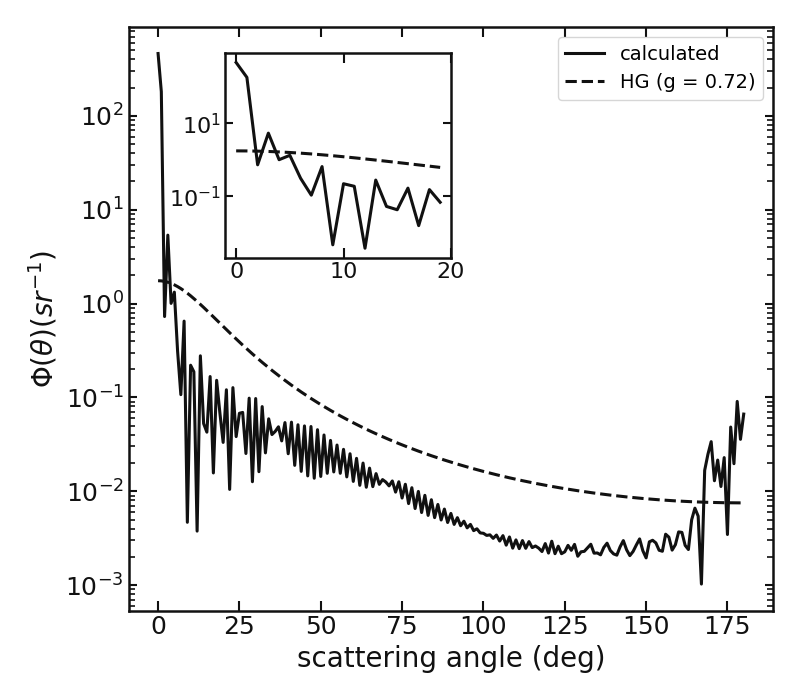}
\caption{Variation of phase function vs scattering angle curves for 10 $\mu m$ size lunar glass dust grains and its comparison with Henyey-Greenstein phase function.}
\label{HGpfcomparison}
\end{figure}

In comparison, the HG phase function has a weak dependence on the scattering angle as the corresponding asymmetric parameter ($g$) is equivalent to isotropic grains, when the size of the grain is small compared to the incident wavelength. As the size of the dust grains increases, the interference features seen in theoretical phase curves is absent in the HG curve as it is a smooth function. The phase function profile of lunarglass in comparison with the HG phase function is shown as an example in the Figure \ref{HGpfcomparison} [see \ref{SRTGconst} for the entire set of comparison plots]. HG phase function has overestimated the values, as compared to the scattering phase function, irrespective of the size and composition of the dust grains. Earlier, the computational discrepancies found in the radiative transfer model using HG phase function during the study of aerosols also give a similar explanation  \citep{1995Marshall,1998Oliver}. The deviation from the scattering phase function is also found when interstellar dust grains are used to model the scattering at the NIR and UV regions \citep{2003Draine,2022Baes}. This signifies the importance of using the theoretical scattering phase function in the radiative transfer models, which is characterised by the size, shape and composition of the dust grains rather than the HG phase function. 


\subsection{Scattered intensity}\label{scatres}

\begin{figure}[h]
    \centering
    \includegraphics[width=0.7\textwidth]{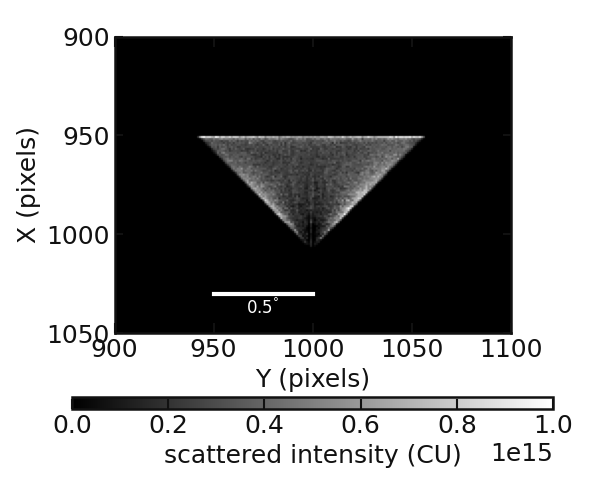}
    \caption{Ejecta curtain scattered intensity map obtained for 0.1 $\mu$m size graphite grains.} 
    \label{scatmap}
\end{figure}

An example of the scattered intensity map of an ejecta curtain for 0.1 $\mu$m size graphite grains is shown in Figure \ref{scatmap}. The median of the obtained model scattered intensity values across the area containing the ejecta curtain in the 2D map and the corresponding median absolute deviation are calculated. 

\subsubsection{Monodisperse grains in the ejecta cone\\}
\begin{figure}[h]
    \centering
    \includegraphics[width=0.7\textwidth]{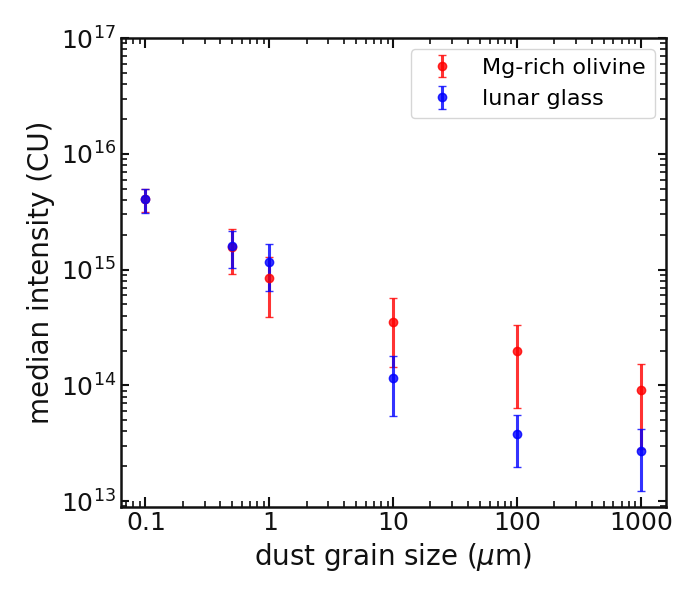}
    \caption{Variation of median scattered intensity in continuum units (1 CU = 1 photons $m^{-2} s^{-1} sr^{-1}$ \AA$^{-1}$) of the ejecta cone with size of the dust grains for different weakly absorbing materials.}
    \label{inonabscomp}
\end{figure}

The variation of the median scattered intensity with the increase of dust grain size is different for different compositions, at a constant number density. The variation of median intensity values for weakly absorbing materials like Mg-rich olivine and lunar glass is shown in the Figure \ref{inonabscomp}. The median scattered intensity is maximum at 0.1 $\mu m$ and then decreases gradually in these materials. The ejecta curtain is brightest at sub-micron-sized dust grains and reduces significantly as the dust grain size reaches the millimetre range. Mg-rich olivine has shown the brightest median scattered intensity at each individual size also. 


\newpage

\begin{figure}[H]
    \centering
    \includegraphics[width=0.7\textwidth]{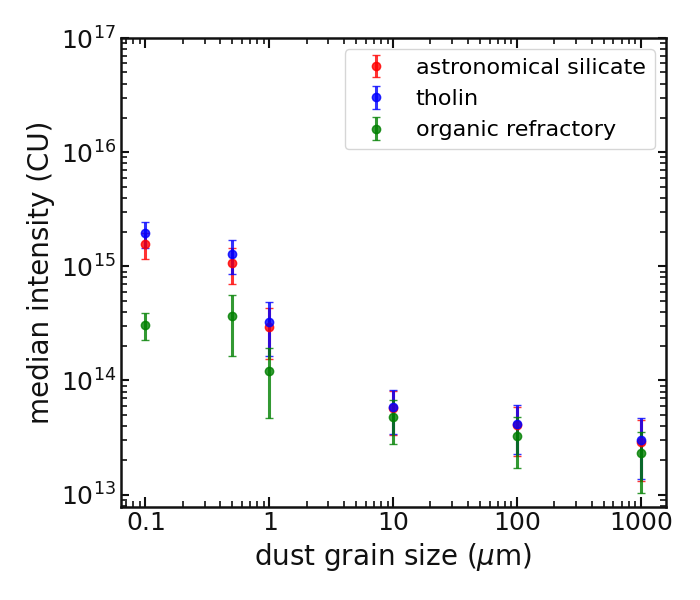}
    \caption{Variation of median scattered intensity (in continuum units) of the ejecta cone with size of the dust grains for moderately absorbing materials.}
    \label{iweakabscomp}
\end{figure}

\newpage

\begin{figure}[H]
    \centering
    \includegraphics[width=0.7\textwidth]{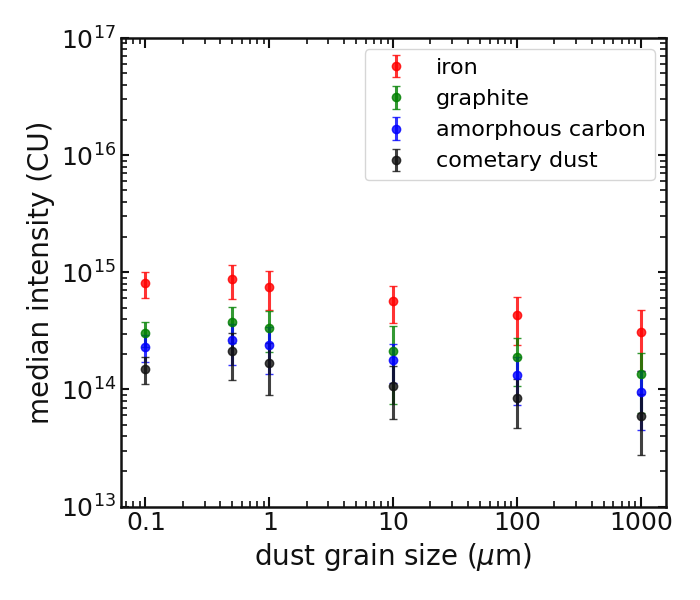}
    \caption{Variation of median scattered intensity (in continuum units) of the ejecta cone with size of the dust grains for strongly absorbing materials.}
    \label{iabscomp}
\end{figure}

For moderately absorbing materials like astronomical silicate, organic refractory and tholin, the median intensity increases, reaches maximum around 0.1 $\mu m$ - 0.5 $\mu m$ then decreases substantially to few order of magnitude (Figure \ref{iweakabscomp}). Similar to the equivalent values of their complex refractive index between astronomical silicate and tholin, the scattered intensity values are also comparable. 
The median scattered intensity variation for absorbing materials is shown in the Figure \ref{iabscomp}. The variation of the median scattered intensity increases, reaches maximum around 0.5 $\mu m$, then decreases slightly for strongly absorbing materials. Cometary dust has shown the least intensity among all the compositions. 

Along with the variation in the median scattered intensity between the compositions, there is a difference in the profile of intensity along the horizontal axis in the scattered intensity map. There is a U-shaped profile seen when dust grains are sub-micron size, and the profile is more flattened as the size increases [as shown in Appendix \ref{SRTGconst}].


\subsubsection{Optical depth dependence in monodisperse grains case\\}\label{optdepthmonodisperse}

\begin{figure}[H]
    \centering
    \includegraphics[width=0.7\textwidth]{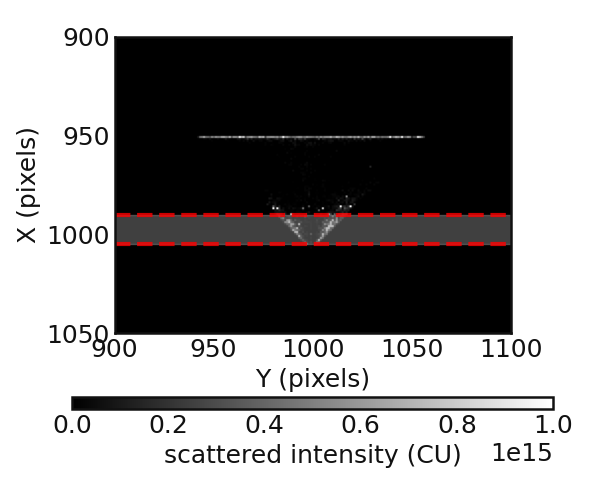}
    \caption{Ejecta curtain scattered intensity map obtained for 0.5 $\mu$m graphite grains when $\tau >$1. The highlighted region is used to calculate the median scattered intensity.}
    \label{scatmapregion}
\end{figure}

Optical depth is one of the major parameters in the study that can affect the scattered intensity results significantly. Here, we have considered three cases of optical depth $\tau <$ 1, $\tau$ = 1 and $\tau >$ 1. The optical depth remains constant throughout the ejecta cone during each individual calculation, but is different for different compositions. The median scattered intensity is calculated from the region as shown in the Figure \ref{scatmapregion}. 
The variation in the median scattered intensity is studied for 0.5 $\mu m$ size grains of all compositions, also varying the dust grain size, keeping the composition  same (eg., amorphous carbon).  


\begin{figure}[H]
    \centering
    \includegraphics[width=0.7\textwidth]{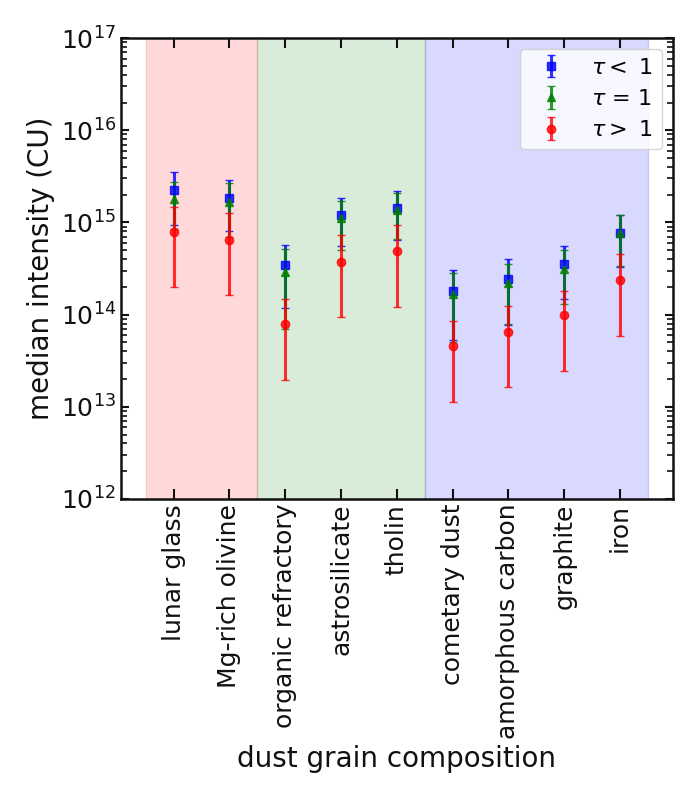}
    \caption{Variation of median scattered intensity for 0.5 $\mu$m sized grains with compositions, at different optical depths. (Here red, green and blue shaded region represents weakly-absorbing, moderately absorbing and strongly absorbing materials respectively.)}
    \label{optdepthplot}
\end{figure}

\begin{figure}[H]
    \centering
    \includegraphics[width=0.7\textwidth]{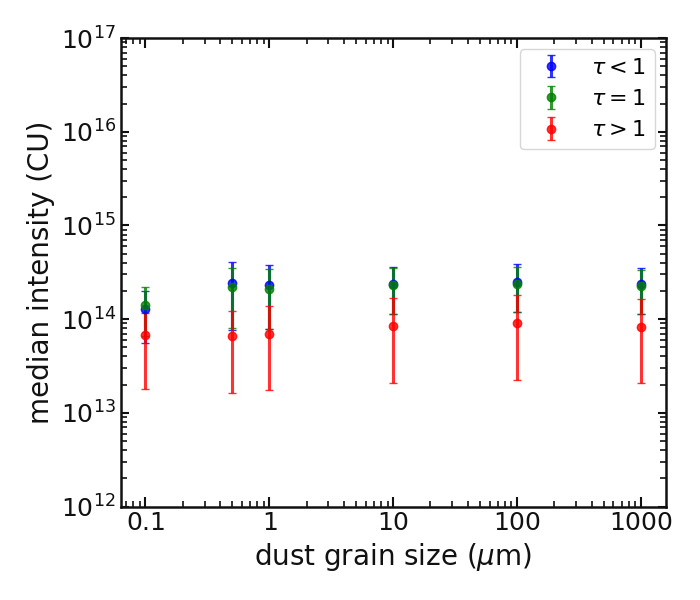}
    \caption{Variation of median scattered intensity (in continuum units) of the ejecta cone with the size of the amorphous carbon dust grains for three different values of optical depth.}
    \label{tauamorcarbon}
\end{figure}

Between the different compositions, keeping the dust grain size same, there is a variation in the median scattered intensity values depending on their absorbing power. We also found a significant decrease with the optical depth along with the variation in the median scattered intensity values, as shown in Figure \ref{optdepthplot}. When the dust grain size is varied, keeping the composition same, the median scattered intensity have remained low for $\tau >$ 1. There is no significant deviation in the intensity profile with the dust grain size and optical depth, as shown in the Figure \ref{tauamorcarbon}. In general, there is a decrease in the magnitude of the median scattered intensity with the increase of optical depth.


\subsubsection{Size distribution in the ejecta cone\\}

In impact events, the dust grain size and number density values are not uniform; rather, they are distributed over a range of values. We have considered two cases with the dust grain-size distribution in the ejecta curtain model, which are: (i) keeping the number density constant throughout the ejecta cone and (ii) varying the number density following a power-law along the axis of the ejecta cone. The dust grains are varied from 0.1 $\mu$m to 1000 $\mu$m following a power-law with the indices -2, -3 and -4. Here, the total number density is kept constant across the power index and compositions, resulting in the different ejecta mass. The ejecta mass and the grain parameters such as \textbf{extinction} cross-sections, scattering cross-sections, single scattering albedo and phase functions are calculated by averaging over the grain-size range.


\begin{figure}[H]
    \centering
    \includegraphics[width=0.7\textwidth]{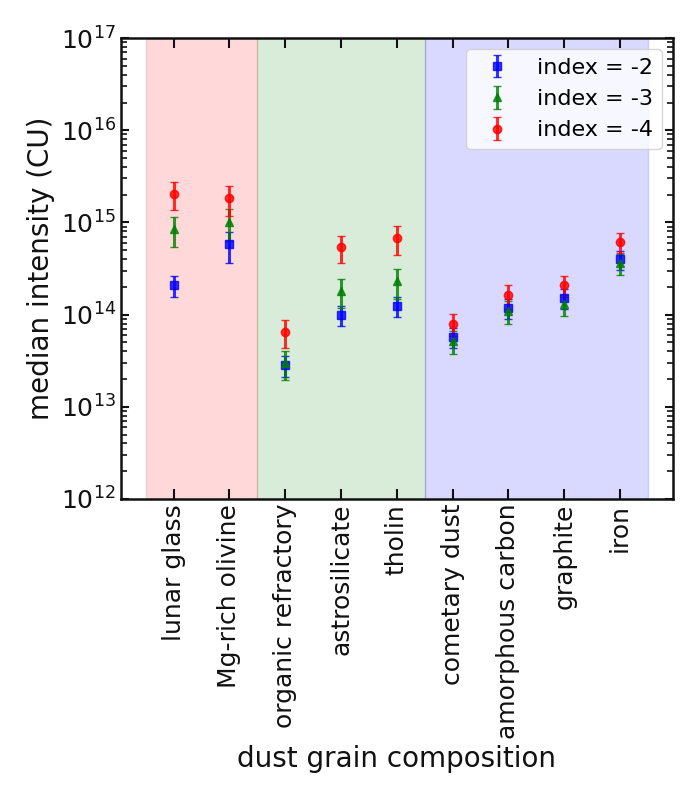}
    \caption{Variation of median scattered intensity (in continuum units) of the ejecta cone with composition by considering a grain-size distribution of different power indices, keeping the constant number density.}
    \label{sizedistributionconstant}
\end{figure}

\begin{figure}[H]
    \centering
    \includegraphics[width=0.7\textwidth]{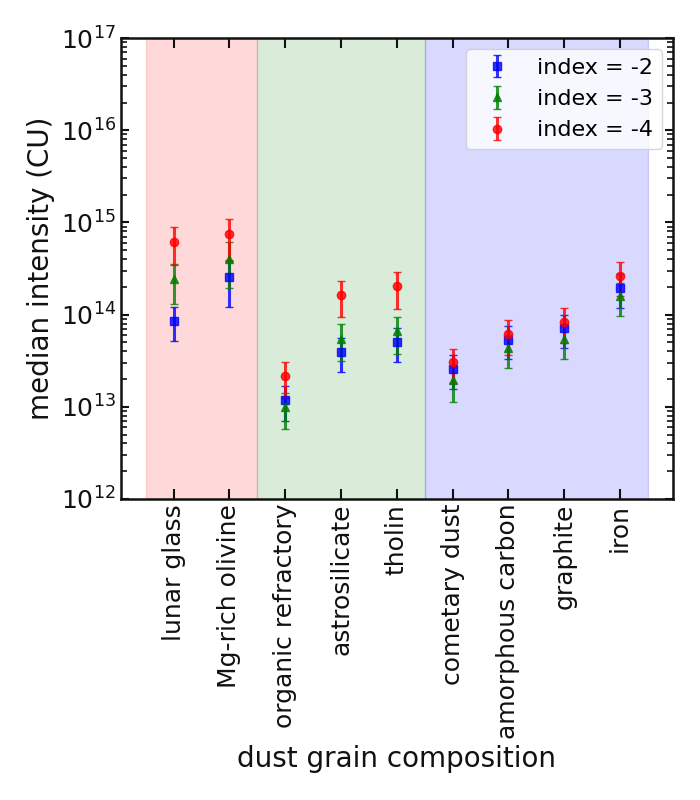}
    \caption{Variation of median scattered intensity (in continuum units) of the ejecta cone with composition by considering a grain-size distribution of different power indices, varying the number density.}
    \label{sizedistributionvar}
\end{figure}

The variation of scattered intensities in each case is shown in the Figure \ref{sizedistributionconstant} and \ref{sizedistributionvar}. It is observed that lower the power index value, higher median scattered intensity, along with the lowest extinction cross-section of the dust grains due to low ejecta mass at constant number density case. Among the different types of absorbing materials, weakly absorbing materials scatters more than others, in each case, independent of ejecta mass and number density. The model intensities mostly follow the pattern of real part of the refractive index and the overall decrease in the median intensity depends on its extinction coefficient between weakly, moderately and strongly absorbing materials. The influence of power index is strong in weakly absorbing material and weak in strongly absorbing material. On the other hand, varying number density results (case ii) in the decrease of the overall median intensities as shown in the Figure \ref{sizedistributionvar}. But the overall behaviour among different compositions and power indices has remained the same.   


\section{Discussion}\label{dis}

The variation of median scattered intensity with the dust grain size is evident from the plots shown in Section \ref{scatres}. The scattered intensity profile across the grain size is expected to be maximum at smaller grain sizes and flattening with the increase in the grain size. The profile is the combined effect of single scattering albedo and extinction cross-section of the dust grains. But the observed reduction in the scattered intensity in multiple scattering scenario at larger grain size is due to the increase in the extinction experienced by the photons after each scattering before reaching the detector. Overall the deviation in the profile of the scattered intensity profile across the grain size, with the special focus on large grain is mostly due to the increase in the extinction cross-section, resulting in an increase in the optical depth.


The presence of absorbing or non-absorbing materials in the excavated materials significantly modifies the scattering intensity, in addition to the variation caused by the dust grain sizes. The weakly-absorbing materials like olivine, silicates which allow constructive interference, enhance the scattering and produce brighter ejecta cone, while strongly absorbing materials like iron dampen the efficiency of the scattering and produces a relatively fainter ejecta cone at sub-micron-sized grains. On the other hand, the absorbing materials like iron-rich, carbonaceous materials showed equivalent brightness, whereas the lunar glass showed the faintest ejecta curtain at larger dust grain sizes. Among the individual compositions, the real part of the complex refractive index also plays a role in enhancing the intensity as seen in amorphous carbon, as well as reduction as seen in cometary dust cases. This indicates that the sub-micron size weakly absorbing dust grains and large carbonaceous dust grains produce the brightest ejecta curtain. 


The scattered intensity depends on the variation of single scattering albedo, scattering phase function and extinction cross-section, where these quantities depend on size and composition of the dust grains. The process not only involves multiple scattering by the dust grains, but also after each scattering, the intensity of each fraction of energy is decreased by the amount of extinction that the photon encounters in the line of sight before it gets detected by the detector. This effect is further tested by increasing the optical depth, which resulted in a decrease in median scattered intensity due to the increase in extinction, as shown in Section \ref{optdepthmonodisperse}. Not only does the extinction term dominate as the size of the dust grains increases, but there is also a sharp decline in the magnitude of the phase function at the scattering angles ($\sim$ 130$^\circ$ - 140$^\circ$), predominantly at larger grain sizes across all the compositions due to forward scattering. The combined effect of both extinction and scattering phase function as detected by the detector gives the uniqueness to the obtained scattering intensity for a fixed size and composition combination.


The inclusion of grain-size distribution and varying number density in the ejecta curtain brings the model close to real impact events. There is a clear distinction between the obtained model intensity as the power index of distribution is varied. The variation in the dust grain size and number density has shown an order difference in the intensity values as compared to the monodisperse grains and uniform optical depth. But the overall profile of intensity variation among different types of absorbing materials has remained the same. The smaller grains dominate at lower index values in the dust grain distribution, which is reflected in the higher scattered intensity values. Further, varying the number density across the ejecta curtain in the model, resulted in a reduction in the magnitude of the intensity, keeping the overall profile the same, due to the increase in ejecta mass and so is optical thickness. Beyond the general understanding, the model also differentiates the different materials depending on their scattering behaviour with the change in the distribution parameters.  

 
Here, we have restricted our calculation to one particular geometry as mentioned in Section \ref{sgeo}, based on Hayabusa2's DCAM3 specifications. This is mainly to reduce the number of variables in the model and to focus the study on the influence of dust parameters on the model intensities. The influence of parameters in Sun-ejecta geometry such as half-opening angle of the cone, incident angle of the solar radiation and height of the ejecta curtain is tested in Appendix \ref{geopartest}. In order to compute approximate values for Ryugu ejecta intensities, we can assume the grains to be 1 mm-sized monodisperse particles made up of silicates. These assumptions are based on the analysis of the DCAM3 images by \citet{2021KojiWada}, which indicated a lack of particles smaller than 1 mm in the ejecta. In addition, the laboratory analysis of the returned samples from Ryugu \citep{2022yadatoru} showed the grains to be similar to CI Chondrites with sizes mostly in the millimetre range. 


However, model predictions which can be compared with observations require the use of accurate ejecta morphology, grain composition, size distribution, grain structure and the relative geometry of the impact event, the Sun and the detector (which determines the phase angle). The calculations are restricted to spherical grains, whereas, in realistic scenarios, the grains are mostly aggregates. The quantitative differences between the intensities are due to differences in the structure of the grain will be addressed in the future. These parameters, apart from being dependent on the impact experiment and the target body, are also time-dependent, and their effect on the observed intensities has to be accounted. Nevertheless, these models can be extremely useful as a preliminary reference for refining scattering models, for extracting grain parameters in impact events and for selecting the target objects for future observations. More importantly, depending on the availability of the observations of past impact events such as DCAM3 observations of Ryugu ejecta, the model can be easily extended to study the time evolution of the ejecta curtain and constrain its parameters.


\section*{List of abbreviations}

\noindent HG - Henyey-Greenstein\\
HRI - High Resolution Instruments\\
LCROSS - Lunar CRater Observation and Sensing Satellite\\
LADEE - Lunar Atmosphere and Dust Environment Explorer\\
DART - Double Asteroid Redirection Test\\
DCAM3 - Deployable Camera 3\\
HST - Hubble Space Telescope\\
IDPs - Interplanetary Dust
Particles\\
SRTG - Scattering Radiative Transfer Grid

\section*{Declarations}

\section*{Availability of data and materials}

The subset of the scattered intensity maps estimated and the Python code to generate the median scattered intensity can be found on zenodo using \ \href{https://doi.org/10.5281/zenodo.22010764}{doi:10.5281/zenodo.22010764}. The additional datasets used and analysed during the current study are available from the corresponding author upon reasonable request.

\section*{Competing interests}

There is no competing interest.

\section*{Funding}

No funding is associated with the present work.

\section*{Authors' contributions}

Keerthana U: visualisation, methodology, investigation, data curation, formal analysis and writing—original draft. 
P. Shalima: conceptualisation, RT modelling, writing—review-editing and supervision.
Koji Wada: conceptualisation, RT modelling, writing—review and editing, and supervision.
Johannes Markkanen: investigation, writing—review and editing. 
Hiroshi Kimura: conceptualisation, RT modelling, writing—review.

\acknowledgments{
KU acknowledges MAHE for providing Dr. TMA Pai Scholarship. We would like to thank the anonymous referees for their insightful and constructive comments, which improved the quality and clarity of our manuscript.
}



\appendix{Convergence test}

\begin{figure}[H]
    \centering
    \includegraphics[width=0.7\textwidth]{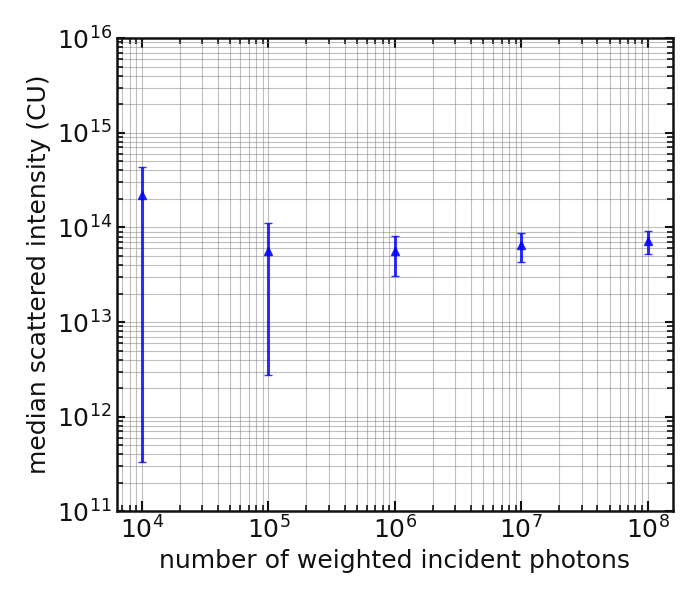}
    \caption{Variation of median scattered intensity (in continuum units) of the ejecta cone with number of weighted photons is tested for organic refractory dust grains following power-law grain size distribution of index -4 with constant number density across the ejecta curtain.}
    \label{convtest}
\end{figure}

The realistic number of incident photons is very large to track individually within the available computational abilities. Thus, we use weighted photons, which is equivalent to a large number of actual incident photons from the Sun. To fix the number of weighted photons used in the model, the convergence test is conducted and is shown in Figure \ref{convtest}. As an example, organic refractory dust grains are considered following a grain size distribution with the power-law index -4, with constant number density across the ejecta curtain. The simulation is repeated for $10^4$ to $10^8$ incident photons. The calculated median intensity across the ejecta curtain nearly stabilised after $10^6$ incident photons. The corresponding median absolute deviation representing the spatial variation across the ejecta curtain is also substantially reduced with the increase of weighted photons. Thus, the weighted incident photons are restricted to $10^7$ photons in the calculations.

\appendix{Dependence of ejecta curtain geometry parameter}\label{geopartest}

\begin{figure}[H]
    \centering
    \includegraphics[width=0.7\textwidth]{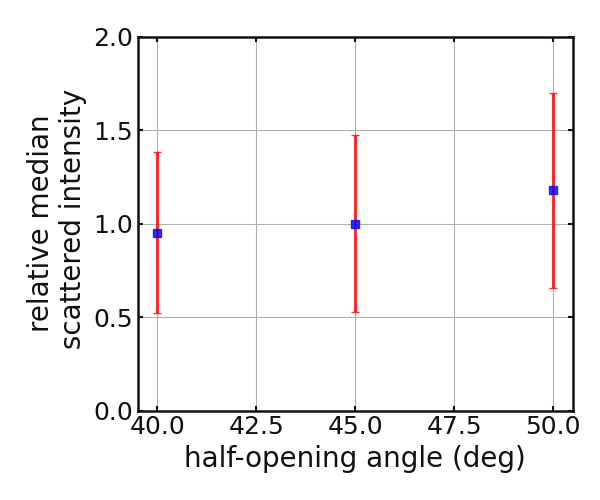}
    \caption{Variation of relative median scattered intensity (in continuum units) of the ejecta cone with half-opening angle of the ejecta cone is tested for organic refractory dust grains following power-law grain size distribution of index -4 with constant number density across the ejecta curtain.}
    \label{eatest}
\end{figure}

\begin{figure}[H]
    \centering
    \includegraphics[width=0.7\textwidth]{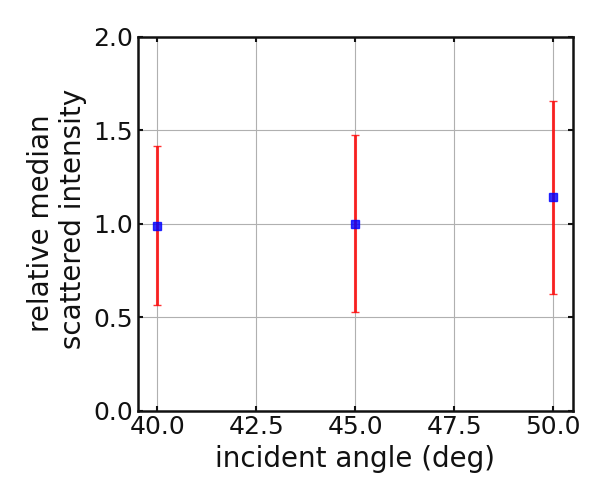}
    \caption{Variation of relative median scattered intensity (in continuum units) of the ejecta cone with incident angle of the solar radiation is tested for organic refractory dust grains following power-law grain size distribution of index -4 with constant number density across the ejecta curtain.}
    \label{iatest}
\end{figure}

\begin{figure}[H]
    \centering
    \includegraphics[width=0.7\textwidth]{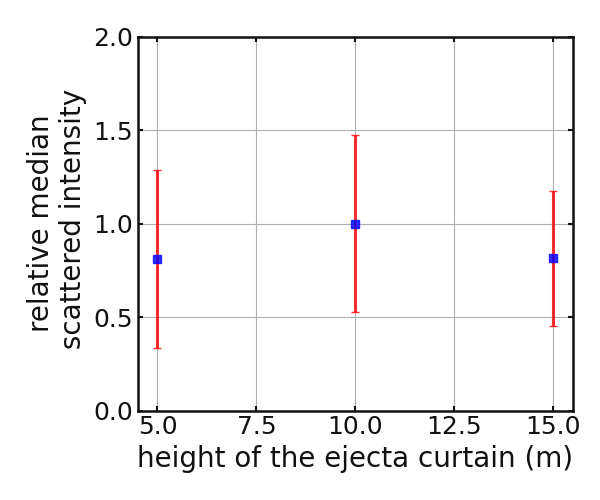}
    \caption{Variation of relative median scattered intensity (in continuum units) of the ejecta cone with height of the ejecta cone is tested for organic refractory dust grains following power-law grain size distribution of index -4 with constant number density across the ejecta curtain.}
    \label{heighttest}
\end{figure}

The parameters involved in the Sun-ejecta geometry that could alter the median scattered intensity are the half-opening angle of the cone, height of the ejecta curtain and the incident angle of the solar radiation. The ideal parameters taken for the study are a 45$^{\circ}$ half-opening angle, a 45$^{\circ}$ incident angle and a 10 m height of the ejecta curtain. But these parameters are different depending on the time of observations after the impact events. Here, the ejecta angle and incident angles are varied as 40$^{\circ}$, 45$^{\circ}$ and 50$^{\circ}$. The height of the ejecta curtain varied as 5 m, 10 m and 15 m. The relative scattered intensities are calculated with respect to the ideal values taken earlier and is shown in the Figures \ref{eatest}, \ref{iatest} and \ref{heighttest}. During the variation of one parameter, all other parameters are kept ideal in the calculations. The error bars represent the median absolute deviation obtained through error propagation. The median scattered intensity has increased with an increase in half-opening angle and incident angle, whereas the median scattered intensity is maximum when the height of the ejecta curtain is 10 m compared to the other values of height considered for the test. The variation in the median intensity with incident angle is mainly due to the influence of the phase function values at the corresponding scattering angles, whereas the variation due to the other parameters are mostly due to the variation in the extinction in the medium.


\appendix{Qualitative comparison of model results with DCAM3 observation}

\begin{figure}[H]
    \centering
    \includegraphics[width=1.0\textwidth]{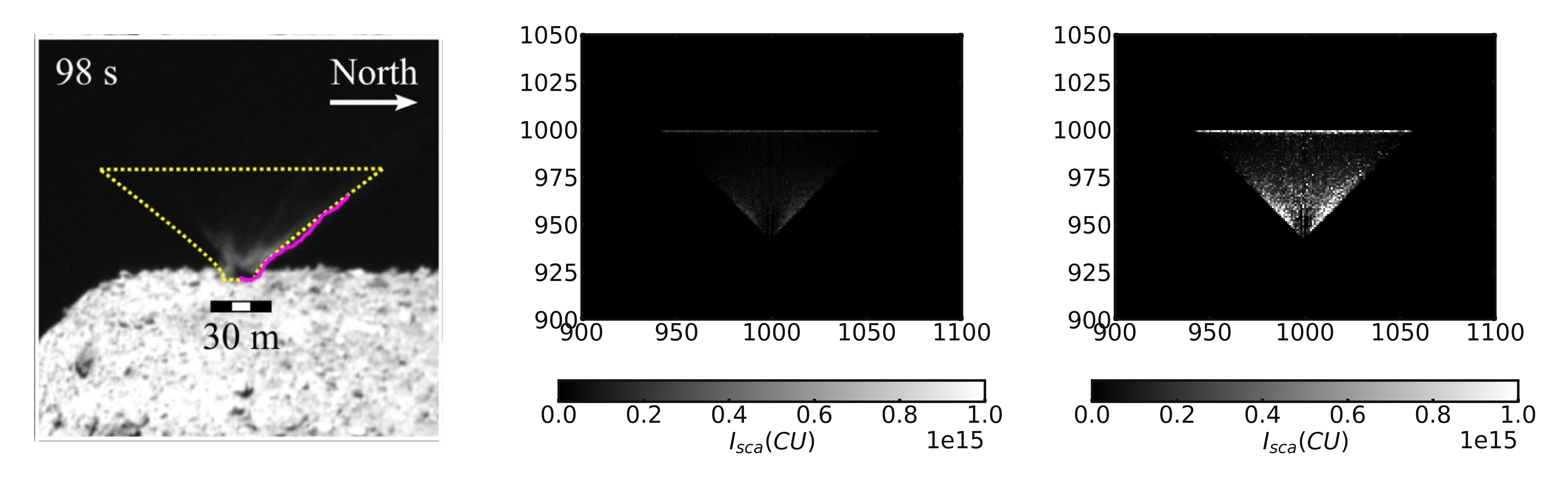}
    \caption{The ejecta curtain captured by DCAM3 at 98 s after the impact experiment (left), 2D scattered intensity map of 1 mm sized astronomical silicate (center) and Mg-rich olivine (right).}
    \label{dcam3comparison}
\end{figure}

The available raw images are captured at different times of the Hayabusa2 mission impact experiment. In order to fix the time of observation to compare the model scattering intensity maps, the dynamical ejecta curtain model given in \citet{2021KojiWada} has been used. Constraining the lower limit of optical depth to 0.001, it is found that the height of the ejecta curtain reached 10 m at 98 s after the impact experiment, considering 1 mm monodisperse ejecta particles across the ejecta curtain. The visual comparison of the scattered maps obtained for 1 mm size astronomical silicate and Mg-rich olivine grains with DCAM3 image at 98s is shown in Figure \ref{dcam3comparison}. Our median model intensities for 1 mm-sized astronomical silicate grains and Magnesium-rich olivine in the ejecta are found to be (0.03 $\pm$ 0.016) $\times 10^{15}$ photons $m^{-2} s^{-1} sr^{-1}$ \AA$^{-1}$ and (0.09 $\pm$ 0.06) $\times 10^{15}$ photons $m^{-2} s^{-1} sr^{-1}$ \AA$^{-1}$ respectively. The detailed comparison considering all the parameters must be investigated in the future. 


\appendix{Scattering radiative transfer grids (SRTG) for monodisperse case and constant number density}\label{SRTGconst}

\begin{figure*}
    \centering
    \includegraphics[scale = 0.23]{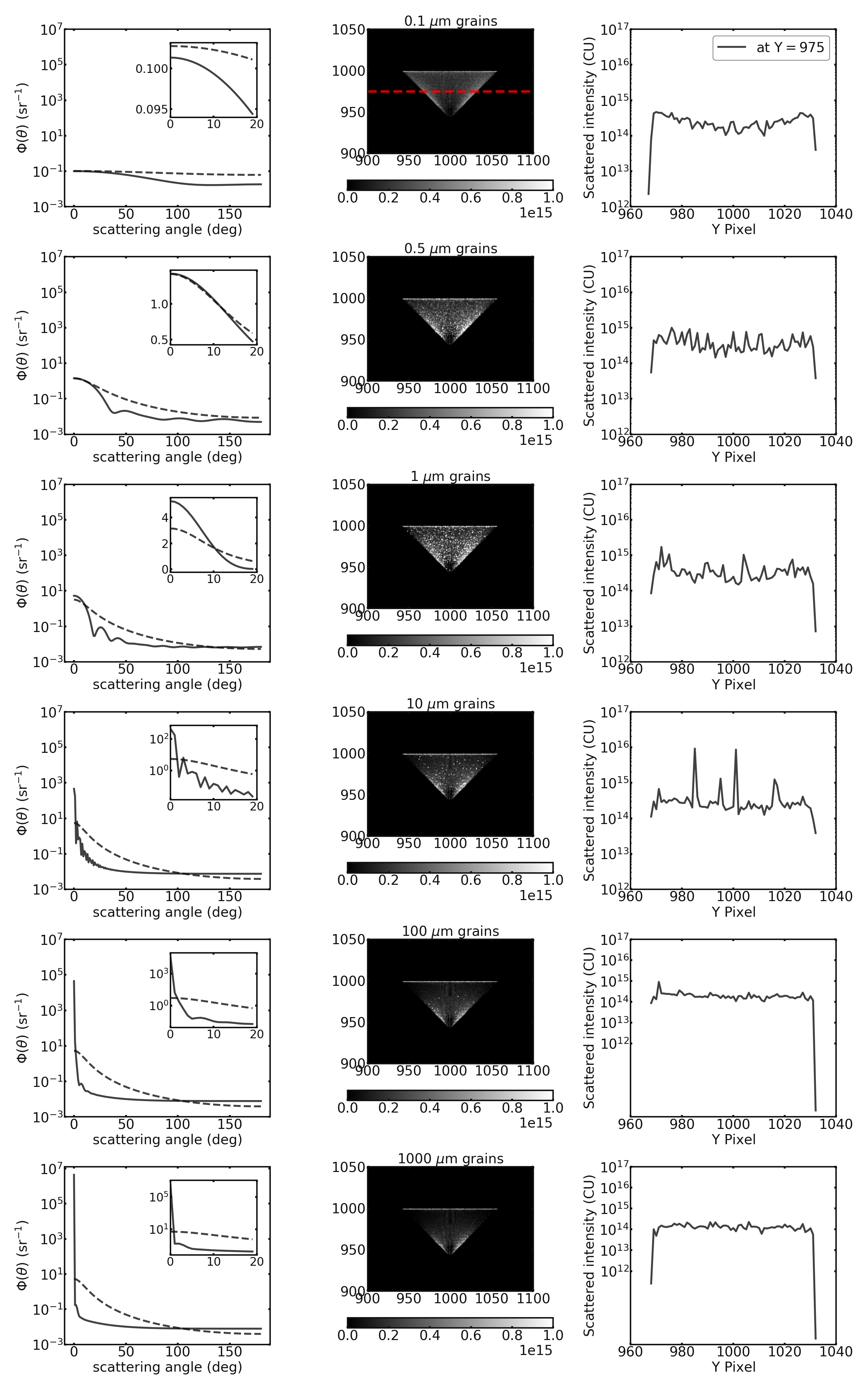}
    \caption{Phase function (left), ejecta curtain 2D intensity map (centre) and the corresponding intensity profile (right) along y-axis at x = 975 for amorphous carbon dust grains of sizes 0.1 $\mu m$ to 1000 $\mu m$.} 
    \label{amorcarbon}
\end{figure*}

\begin{figure*}
    \centering
    \includegraphics[scale = 0.23]{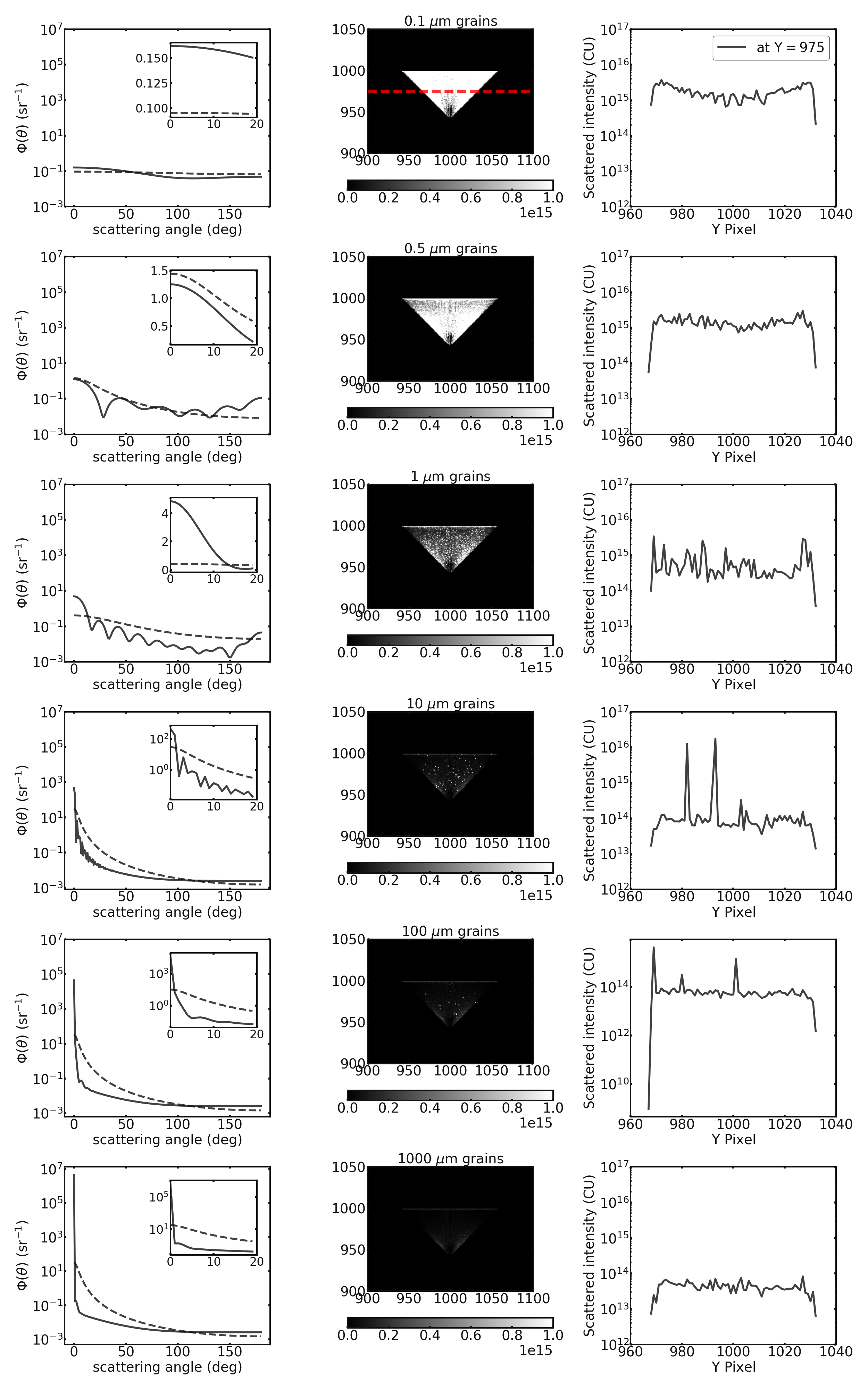}
    \caption{Phase function (left), ejecta curtain 2D intensity map (centre) and the corresponding intensity profile (right) along y-axis at x = 975 for astronomical silicate dust grains of sizes 0.1 $\mu m$ to 1000 $\mu m$.}
    \label{astrosil}
\end{figure*}

\begin{figure*}
    \centering
    \includegraphics[scale = 0.23]{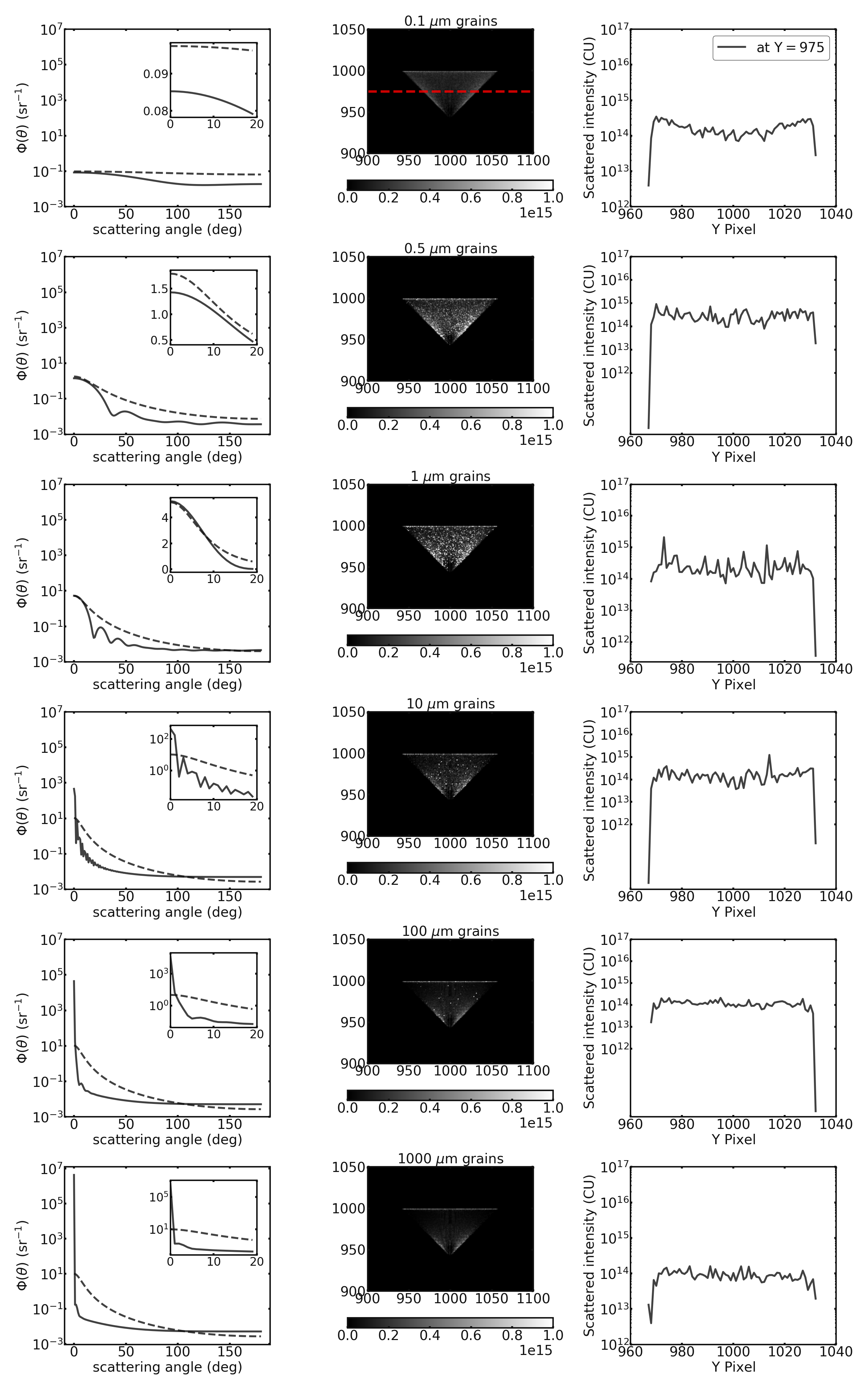}
    \caption{Phase function (left), ejecta curtain 2D intensity map (centre) and the corresponding intensity profile (right)  along y-axis at x = 975 for cometary dust grains of sizes 0.1 $\mu m$ to 1000 $\mu m$.}
    \label{comdust}
\end{figure*}

\begin{figure*}
    \centering
    \includegraphics[scale = 0.23]{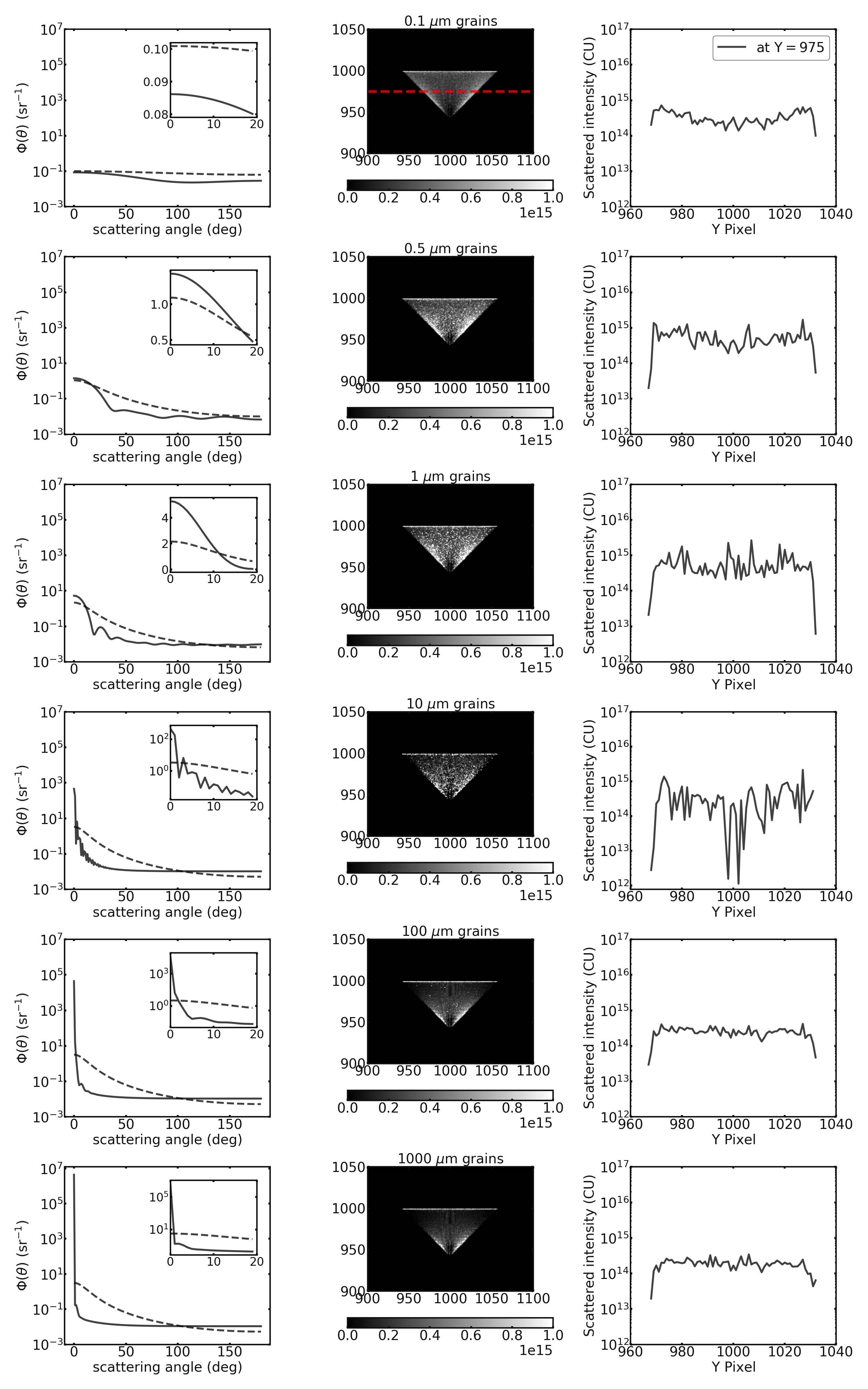}
    \caption{Phase function (left), ejecta curtain 2D intensity map (centre) and the corresponding intensity profile (right) along y-axis at x = 975 for graphite dust grains of sizes 0.1 $\mu m$ to 1000 $\mu m$.}
    \label{graphite}
\end{figure*}

\begin{figure*}
    \centering
    \includegraphics[scale = 0.23]{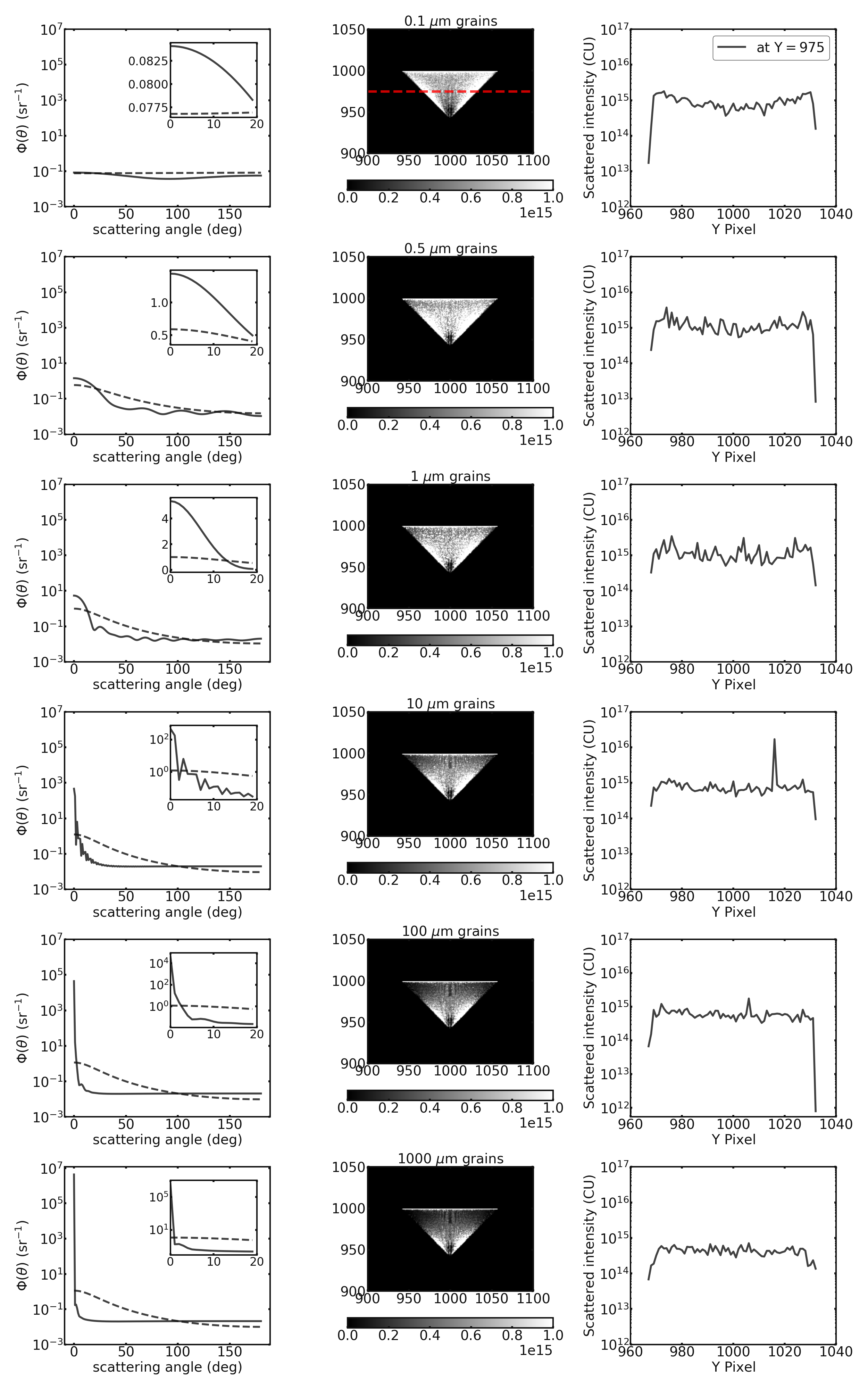}
    \caption{Phase function (left), ejecta curtain 2D intensity map (centre) and the corresponding intensity profile (right) along y-axis at x = 975 for iron dust grains of sizes 0.1 $\mu m$ to 1000 $\mu m$.}
    \label{iron}
\end{figure*}

\begin{figure*}
    \centering
    \includegraphics[scale = 0.23]{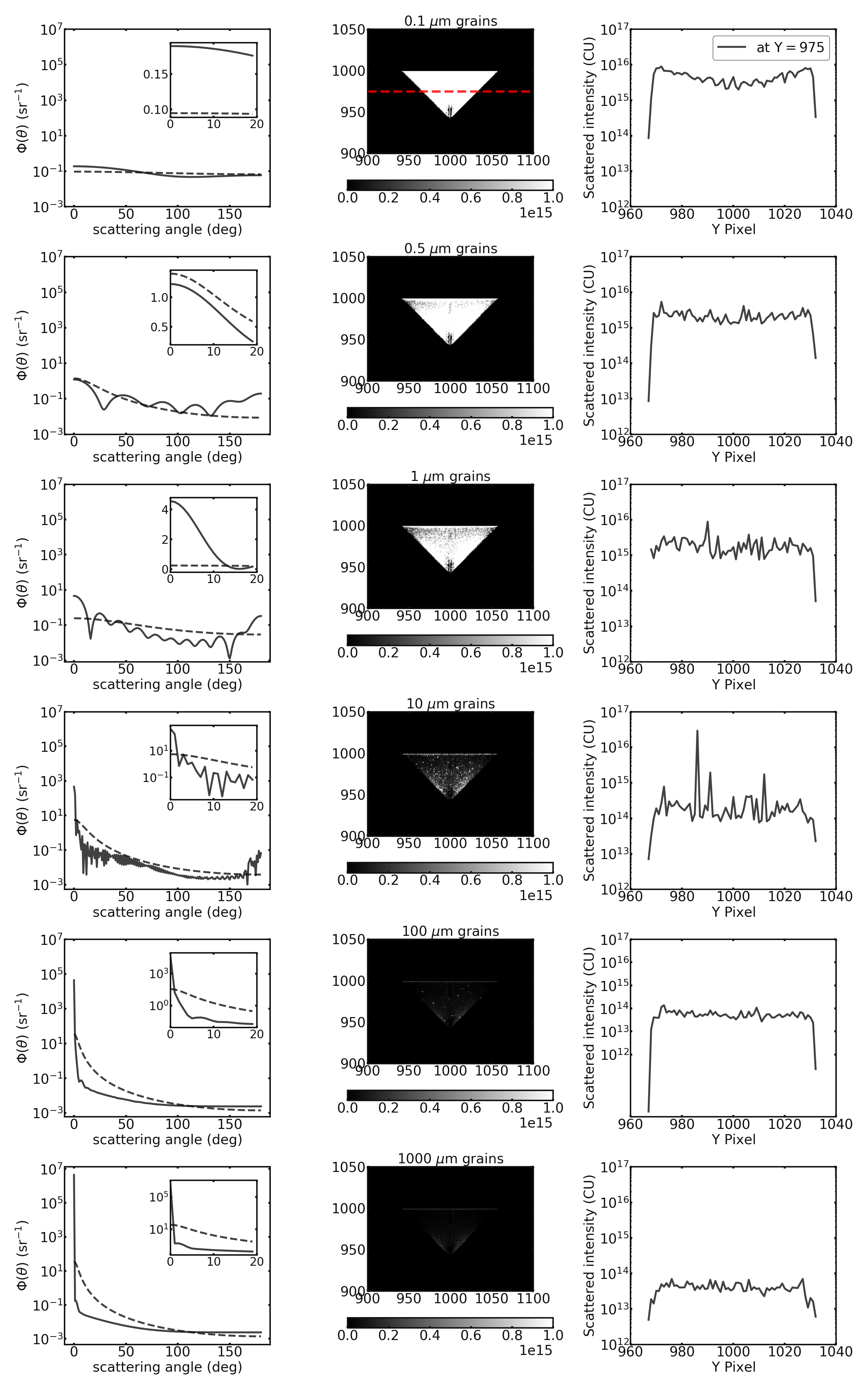}
    \caption{Phase function (left), ejecta curtain 2D intensity map (centre) and the corresponding intensity profile (right) along y-axis at x = 975 for lunar glass dust grains of sizes 0.1 $\mu m$ to 1000 $\mu m$.}
    \label{lunarglass}
\end{figure*}

\begin{figure*}
    \centering
    \includegraphics[scale = 0.23]{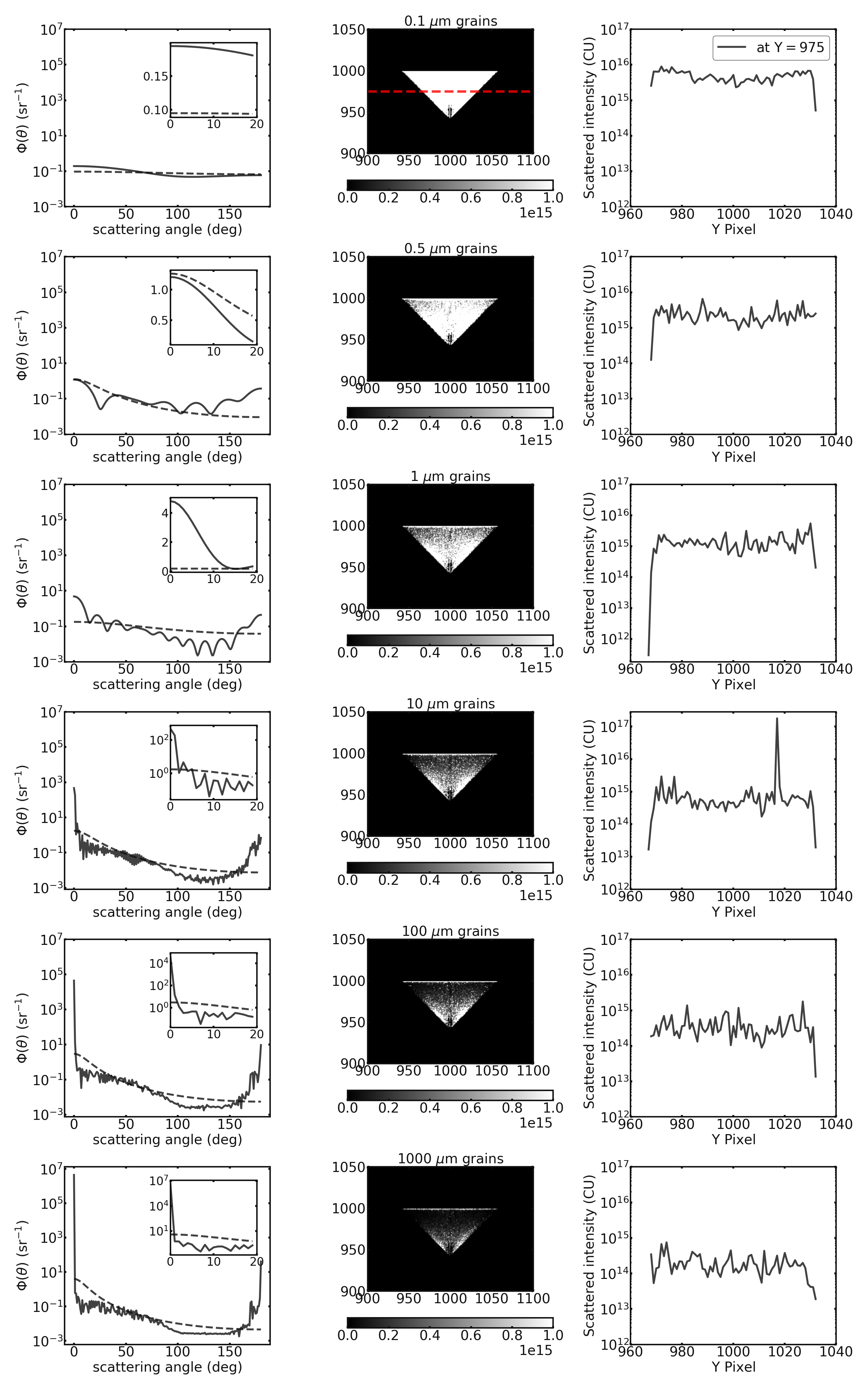}
    \caption{Phase function (left), ejecta curtain 2D intensity map (centre) and the corresponding intensity profile (right) along y-axis at x = 975 for magnesium-rich olivine dust grains of sizes 0.1 $\mu m$ to 1000 $\mu m$.}
    \label{olivine}
\end{figure*}

\begin{figure*}
    \centering
    \includegraphics[scale = 0.23]{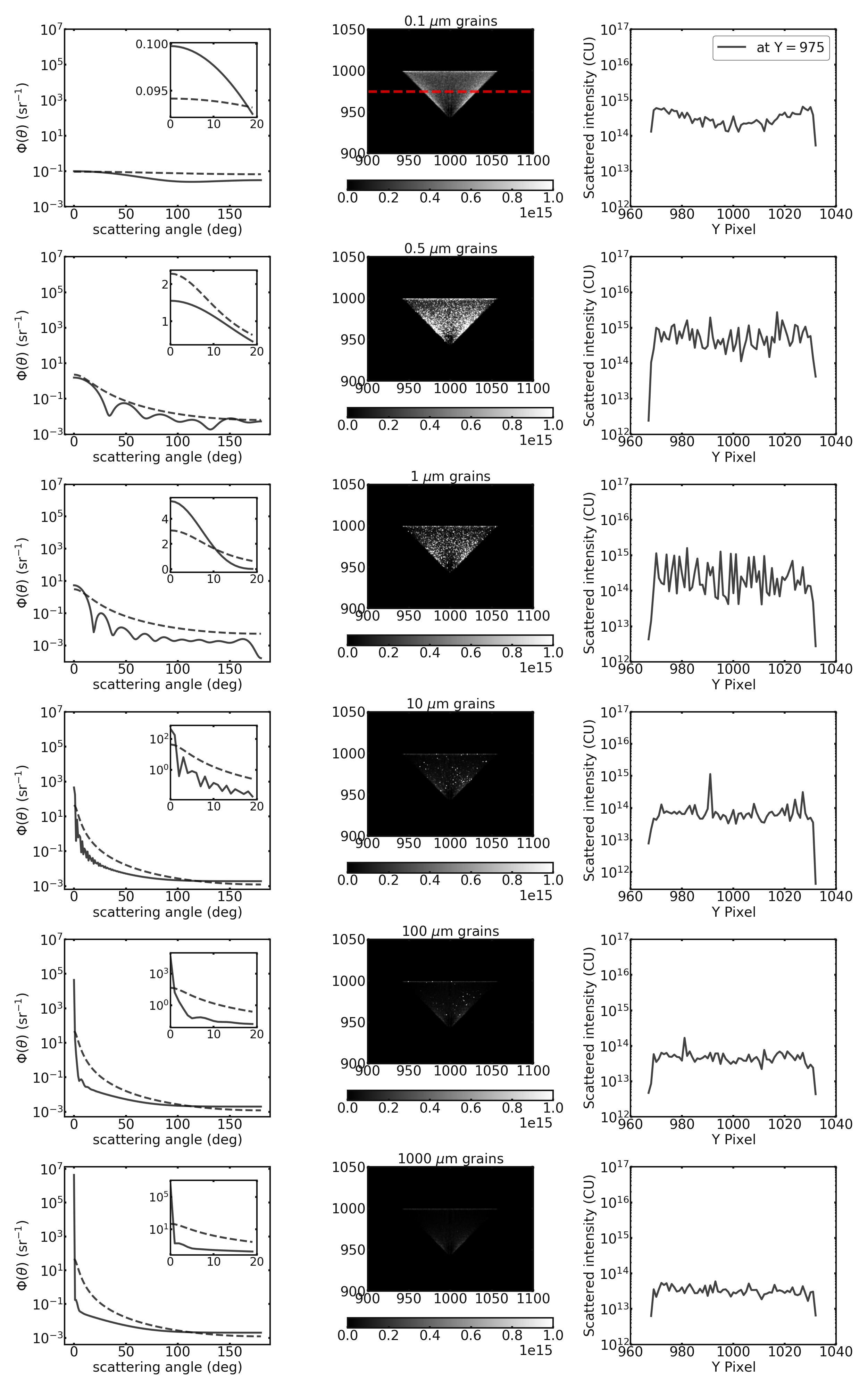}
    \caption{Phase function (left), ejecta curtain 2D intensity map (centre) and the corresponding intensity profile (right) along y-axis at x = 975 for organic refractory dust grains of sizes 0.1 $\mu m$ to 1000 $\mu m$.}
    \label{orgrefra}
\end{figure*}

\begin{figure*}
    \centering
    \includegraphics[scale = 0.23]{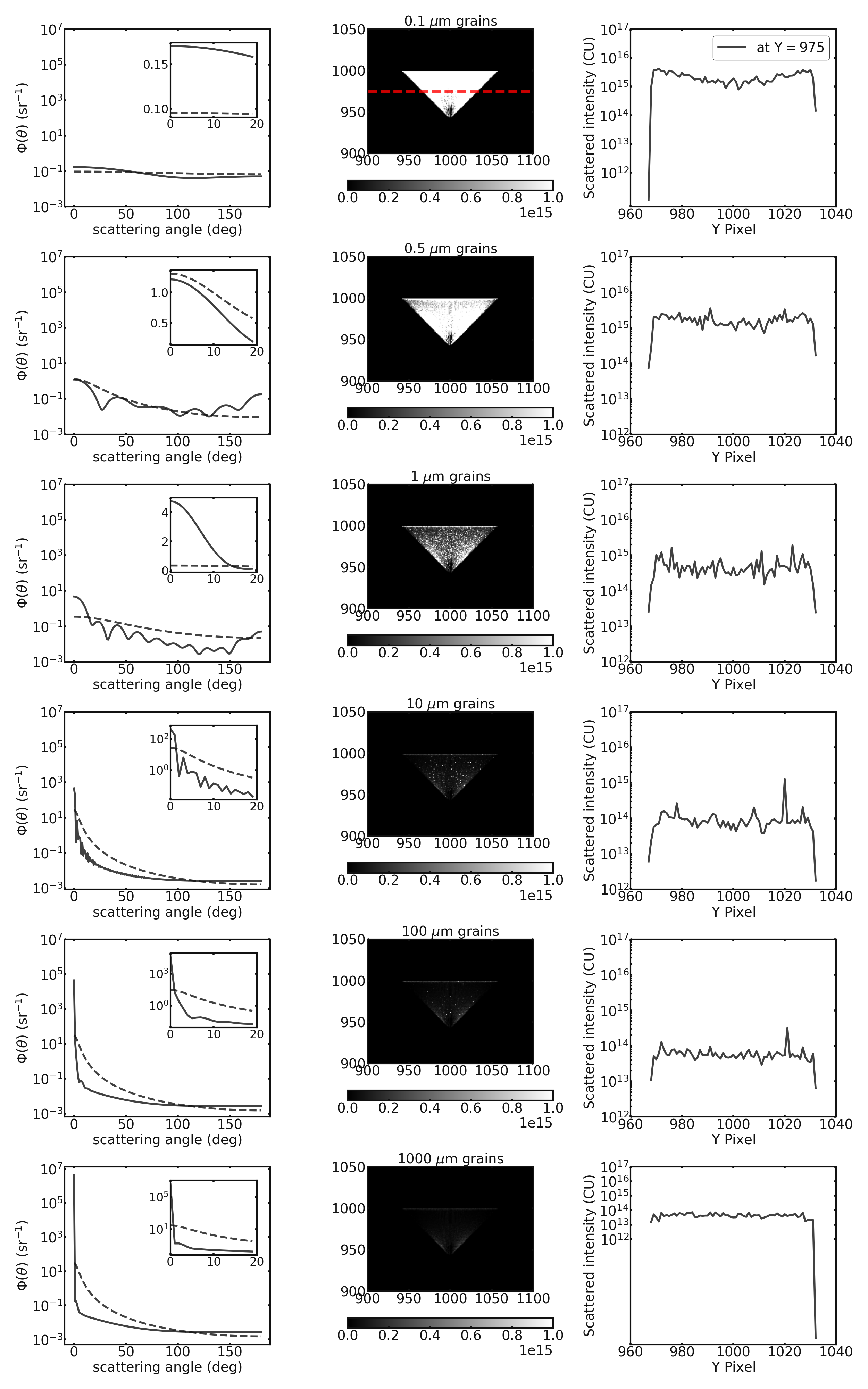}
    \caption{Phase function (left), ejecta curtain 2D intensity map (centre) and the corresponding intensity profile (right)  along y-axis at x = 975 for tholin dust grains of sizes 0.1 $\mu m$ to 1000 $\mu m$.}
    \label{tholin}
\end{figure*}

The results are presented as a SRTG. In the given SRTG, the three images horizontally represent
\begin{enumerate}
    \item A plot of phase function $\Phi(\theta)$ as a function of scattering angle and its comparison with the Henyey-Greenstein (HG) phase function.
    \item 2D intensity map of an ejecta curtain.
    \item A plot showing the variation of model intensity along the x-axis at y = 975 in the ejecta curtain map.
\end{enumerate}
\noindent The images vertically represent the variation of the scattering phase function, ejecta curtain intensity map, and scattered intensity profiles with the dust grain size ranging from 0.05 $\mu m$ to 1000 $\mu m$. The grid of listed compositions is given in the Figures \ref{amorcarbon}, \ref{astrosil}, \ref{comdust}, \ref{graphite}, \ref{iron}, \ref{lunarglass}, \ref{olivine}, \ref{orgrefra} and \ref{tholin}.  

\begin{table}[H]
\centering
\caption{Asymmetry parameter for a given set of compositions and sizes of the dust grain used for comparison.\label{gparameter}}
\begin{tabular}{l c c c c} 
\hline
\hline
Composition & \multicolumn{4}{c}{asymmetry parameter}\\
\hline
radius ($\mu m$) & 0.05 $\mu m$ & 0.1 $\mu m$ & 0.5 $\mu m$  & 1 $\mu m$\\
\hline
Amorphous carbon & 0.0207 & 0.0838 & 0.6914 & 0.7874\\
Astronomical silicate & 0.0147 & 0.0583 & 0.6942 & 0.4662\\
Cometary dust & 0.0166 & 0.0664 & 0.7230 & 0.8322\\
Graphite & 0.0205 & 0.0780 & 0.6527 & 0.7462\\
Iron & 0.0019 & -0.0119 & 0.5437 & 0.6371 \\
Lunar glass & 0.0145 & 0.0574 & 0.6898 & 0.3403\\
Magnesium-rich Olivine & 0.0147 & 0.0582 & 0.6745 & 0.2511\\
Organic refractory & 0.0139 & 0.0554 & 0.7529 & 0.7863\\
Tholin & 0.0148 & 0.0588 & 0.6801 & 0.4290 \\
\hline
\hline
 radius ($\mu m$) & 10 $\mu m$ & 100 $\mu m$ & 1000 $\mu m$ \\
\hline
Amorphous carbon & 0.8362 & 0.8347 & 0.8331\\
Astronomical silicate & 0.9286 & 0.9311 & 0.9306\\
Cometary dust & 0.8801 & 0.8798 & 0.8787\\
Graphite & 0.7921 & 0.7887 & 0.7866\\
Iron & 0.6716 & 0.6614 & 0.6576\\
Lunar glass & 0.8379 & 0.9340 & 0.9342 \\
Magnesium-rich Olivine & 0.7203 & 0.7802 & 0.8095\\
Organic refractory & 0.9411 & 0.9428 & 0.9424\\
Tholin & 0.9248 & 0.9288 & 0.9282\\
\hline
\end{tabular}
\end{table}


\appendix{SRTG considering dust grain-size distribution, number density distribution}\label{SRTGvar}

\begin{figure}[H]
    \centering
    \includegraphics[scale = 0.23]{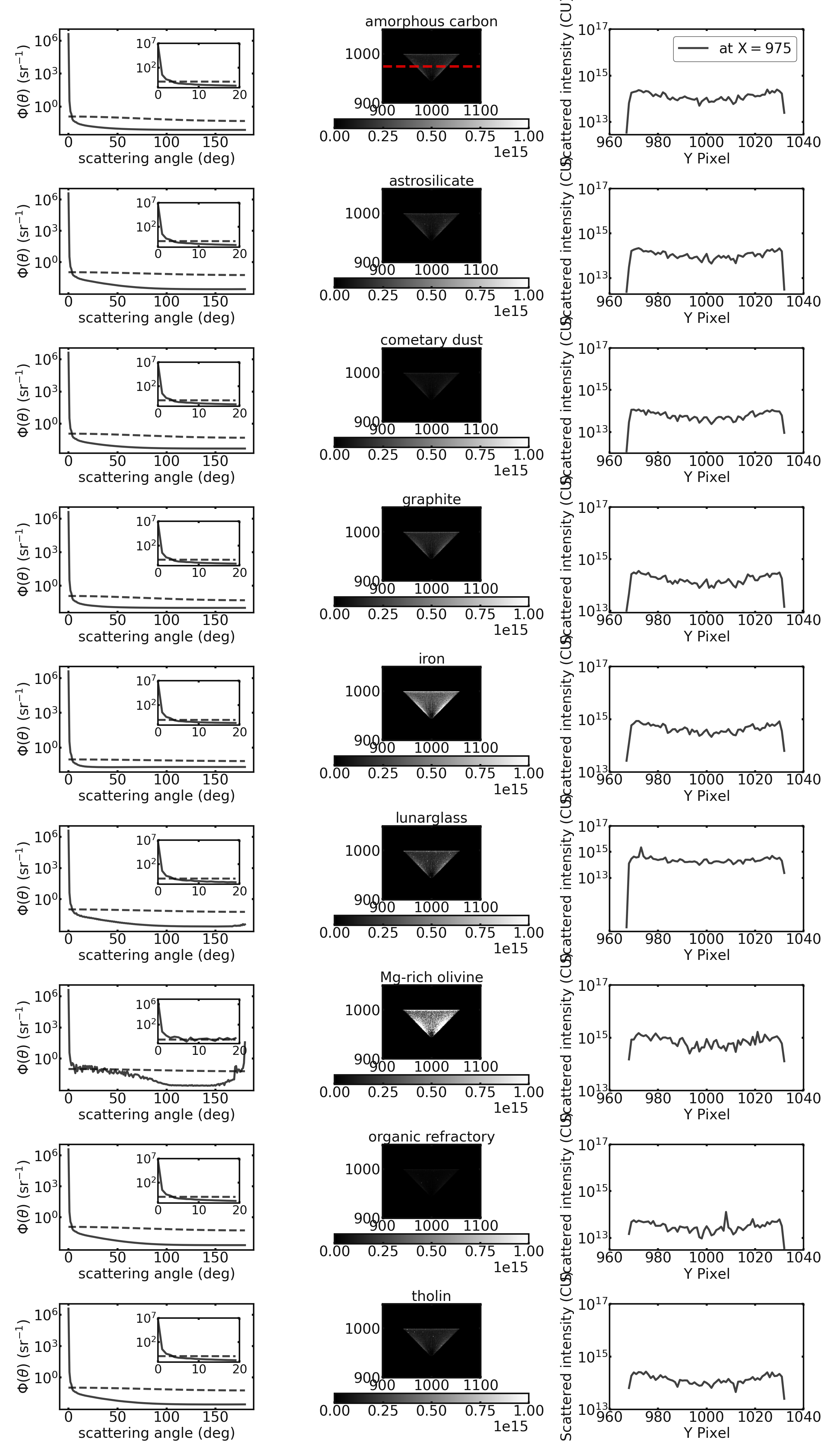}
    \caption{Phase function (left), ejecta curtain 2D intensity map (centre) and the corresponding intensity profile (right)  along y-axis at x = 975 for different compositions considering the size distribution with power-index -2 and constant number density in the ejecta cone.}
    \label{compgridind2}
\end{figure}

\begin{figure}[H]
    \centering
    \includegraphics[scale = 0.23]{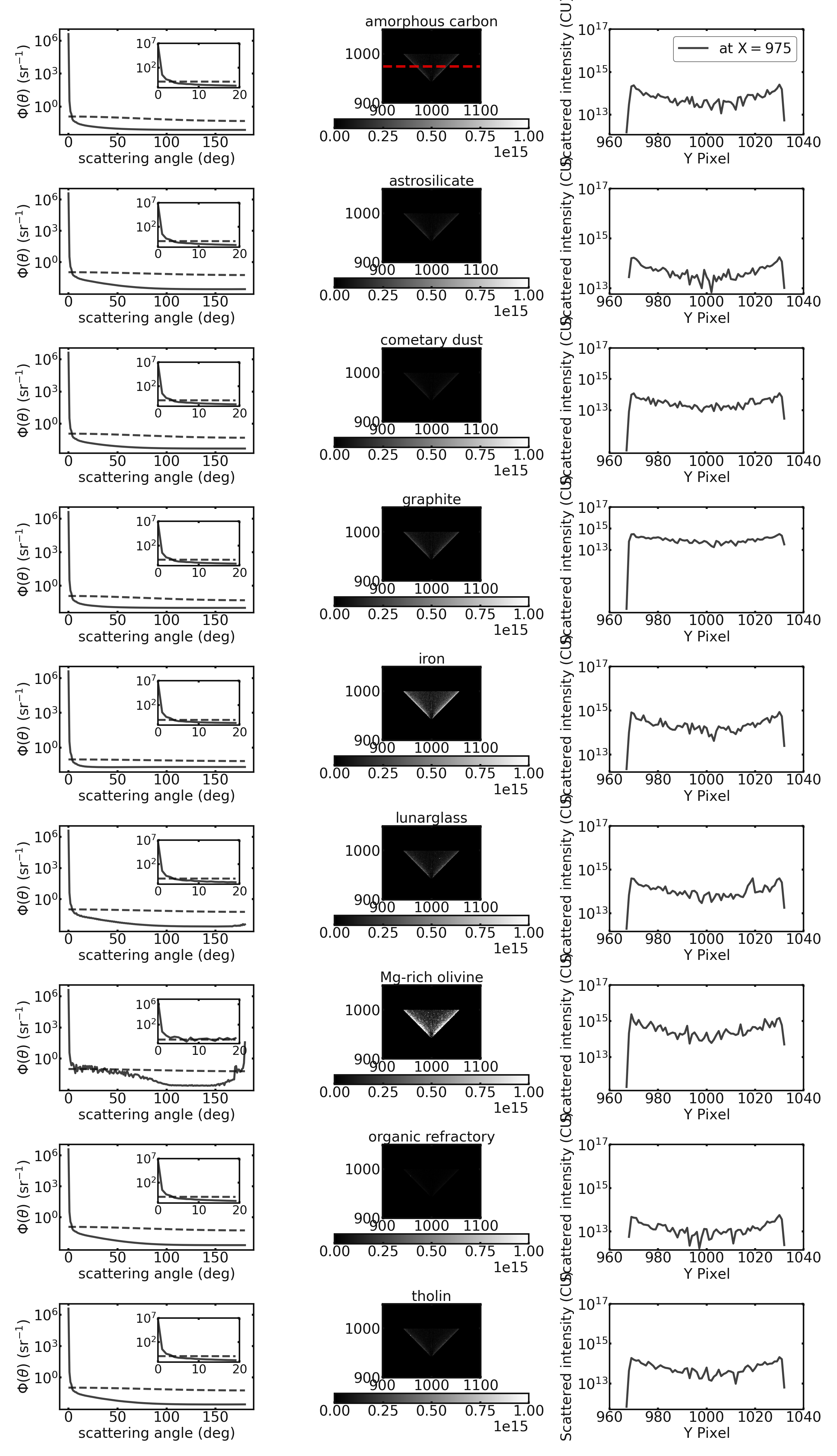}
    \caption{Phase function (left), ejecta curtain 2D intensity map (centre) and the corresponding intensity profile (right)  along y-axis at x = 975 for different compositions considering the size distribution with power-index -2 and varying number density distribution in the ejecta cone.}
    \label{compvargridind2}
\end{figure}

\begin{figure}[H]
    \centering
    \includegraphics[scale = 0.23]{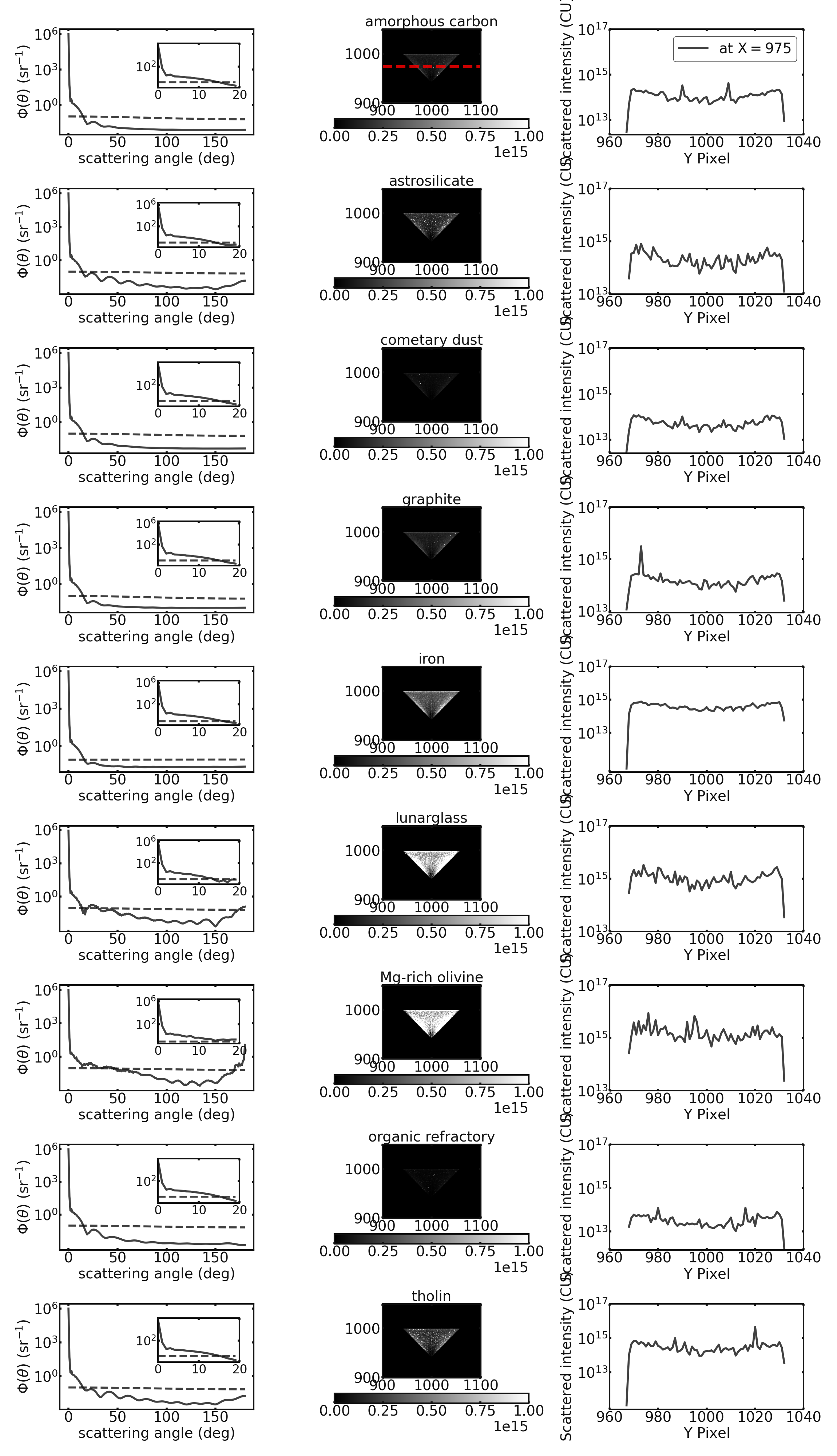}
    \caption{Phase function (left), ejecta curtain 2D intensity map (centre) and the corresponding intensity profile (right)  along y-axis at x = 975 for different compositions considering the size distribution with power-index -3 and constant number density in the ejecta cone.}
    \label{compgridind3}
\end{figure}

\begin{figure}[H]
    \centering
    \includegraphics[scale = 0.23]{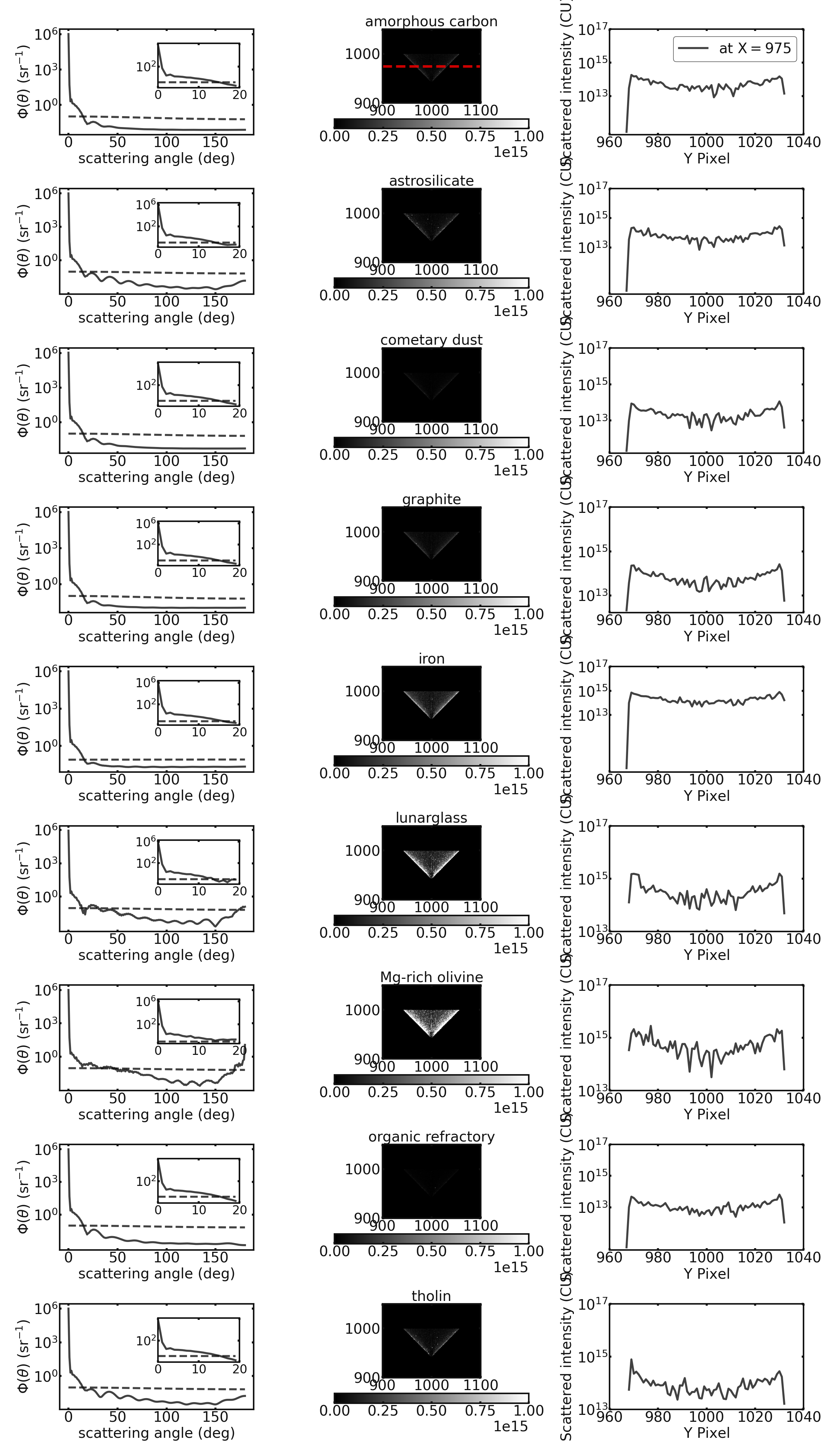}
    \caption{Phase function (left), ejecta curtain 2D intensity map (centre) and the corresponding intensity profile (right)  along y-axis at x = 975 for different compositions considering the size distribution with power-index -3 and varying number density distribution in the ejecta cone.}
    \label{compvargridind3}
\end{figure}

\begin{figure}[H]
    \centering
    \includegraphics[scale = 0.23]{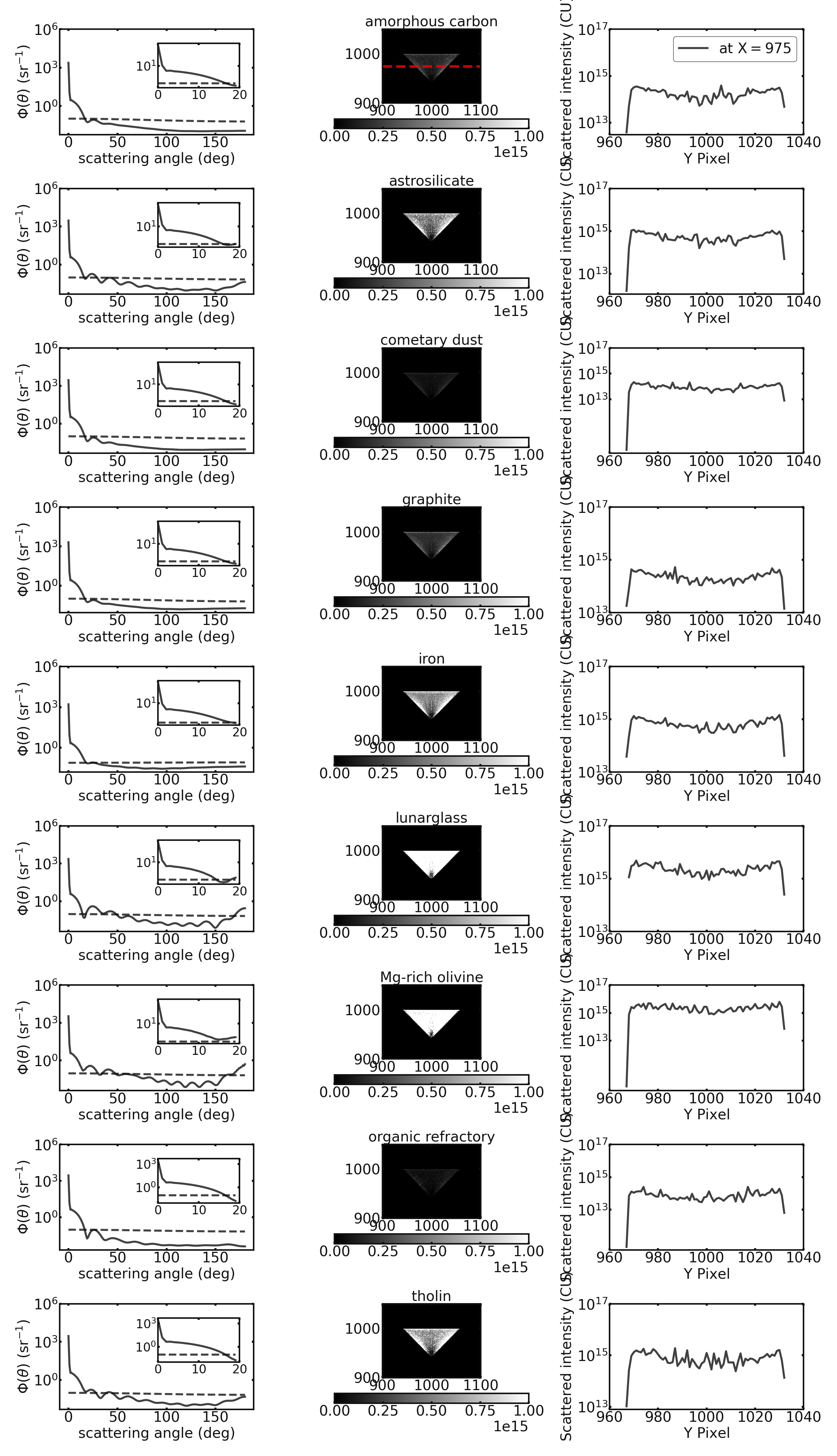}
    \caption{Phase function (left), ejecta curtain 2D intensity map (centre) and the corresponding intensity profile (right)  along y-axis at x = 975 for different compositions considering the size distribution with power-index -4 and constant number density in the ejecta cone.}
    \label{compgridind4}
\end{figure}

\begin{figure}[H]
    \centering
    \includegraphics[scale = 0.23]{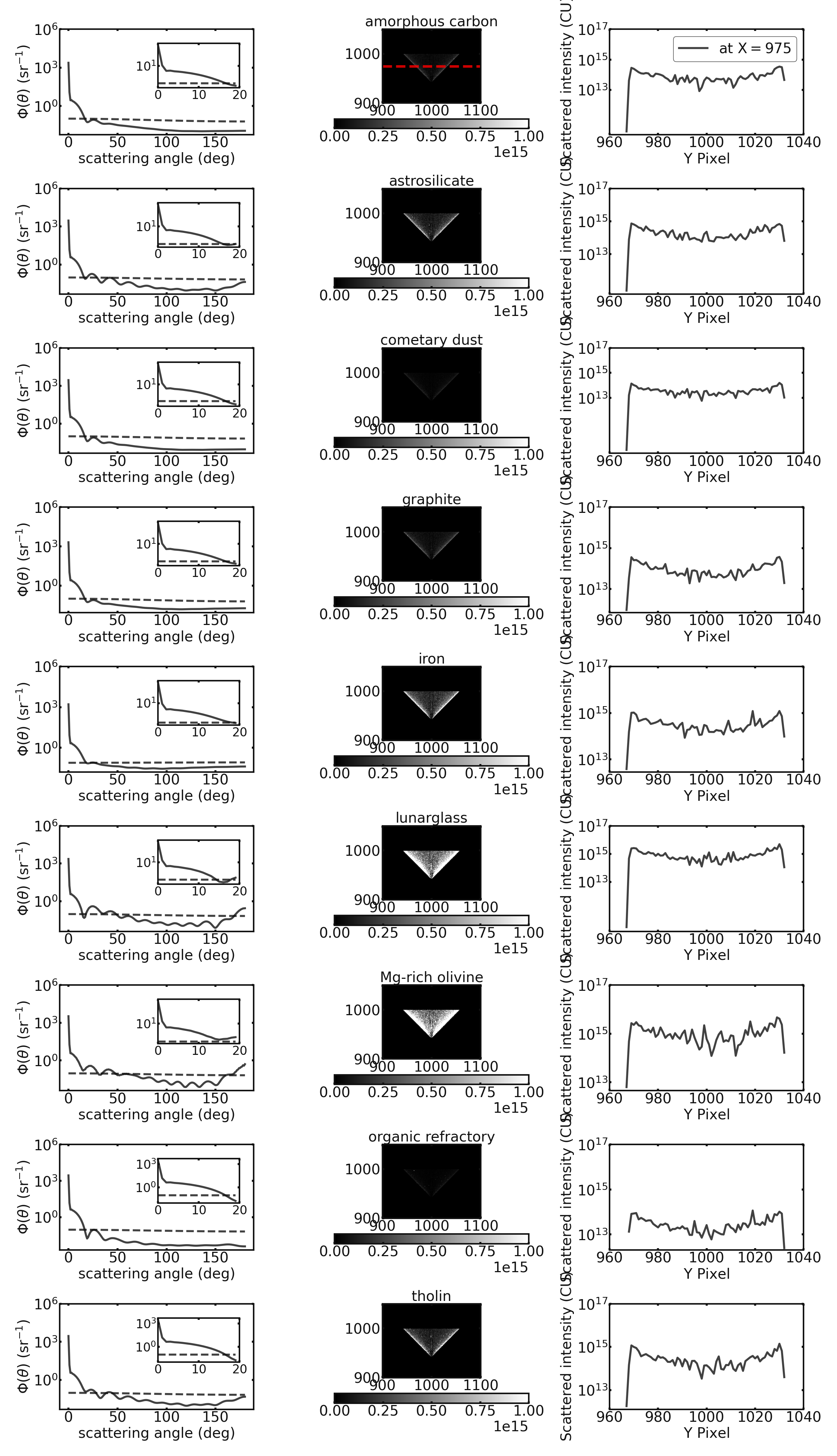}
    \caption{Phase function (left), ejecta curtain 2D intensity map (centre) and the corresponding intensity profile (right)  along y-axis at x = 975 for different compositions considering the size distribution with power-index -4 and varying number density distribution in the ejecta cone.}
    \label{compvargridind4}
\end{figure}

Similar to the previous section, the images vertically represent the variation of the scattering phase function, ejecta curtain intensity map, and scattered intensity profiles for different compositions considering dust grain-size distribution with power indices -2, -3, -4, keeping number density constant as well as following power-law distribution. The grid of listed compositions are given in the Figures \ref{compgridind2}, \ref{compvargridind2}, \ref{compgridind3}, \ref{compvargridind3}, \ref{compgridind4} and \ref{compvargridind4}. 

\begin{table}[H]
\centering
\caption{Asymmetry parameter for a given set of compositions and sizes of the dust grain used for comparison.\label{gparameter_sizedistr}}
\begin{tabular}{l c c c } 
\hline
\hline
Composition & \multicolumn{3}{c}{asymmetry parameter}\\
\hline
power-index & 2 & 3 & 4\\
\hline
Amorphous carbon & 0.1547 &  0.0909 &  0.08457\\
Astronomical silicate & 0.1037 & 0.0624 & 0.05873\\
Cometary dust & 0.1434 & 0.07408 & 0.06718\\
Graphite & 0.1452 & 0.08469 & 0.07867\\
Iron & 0.0533 & -0.0054 & -0.01126\\
Lunar glass & 0.09076 & 0.06028 & 0.05768\\
Magnesium-rich Olivine & 0.0822 & 0.060209 & 0.05842\\
Organic refractory & 0.1301 & 0.0628 & 0.05620 \\
Tholin & 0.1008 & 0.06263 & 0.05925 \\
\hline
\end{tabular}
\end{table}


\appendix{Variation of ejecta mass with dust grain size and composition}

\begin{figure}[H]
    \centering
    \includegraphics[width=0.7\textwidth]{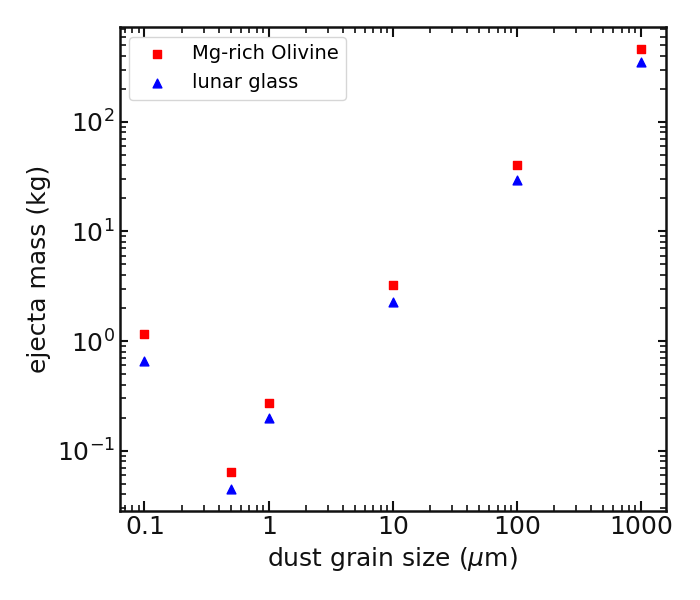}
    \caption{Variation of ejecta mass with dust grain size for weakly absorbing materials.}
    \label{mej_weak}
\end{figure}

\begin{figure}[H]
    \centering
    \includegraphics[width=0.7\textwidth]{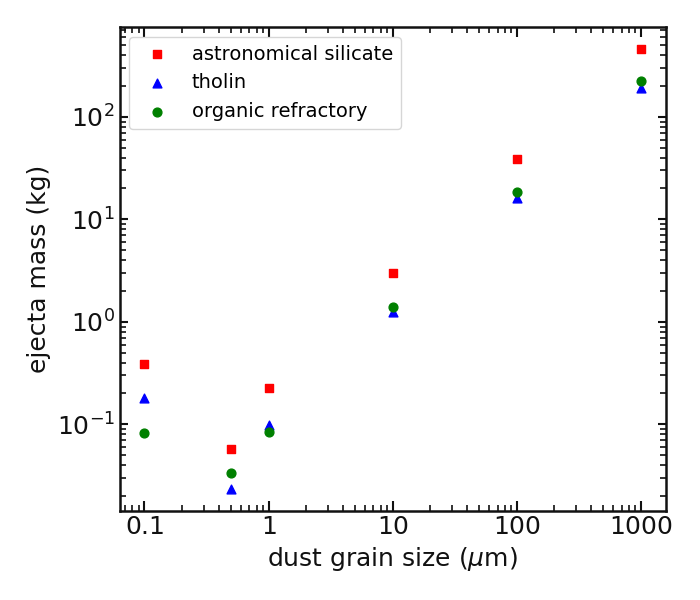}
    \caption{Variation of ejecta mass with dust grain size for moderately absorbing materials}
    \label{mej_moderate}
\end{figure}

\begin{figure}[H]
    \centering
    \includegraphics[width=0.7\textwidth]{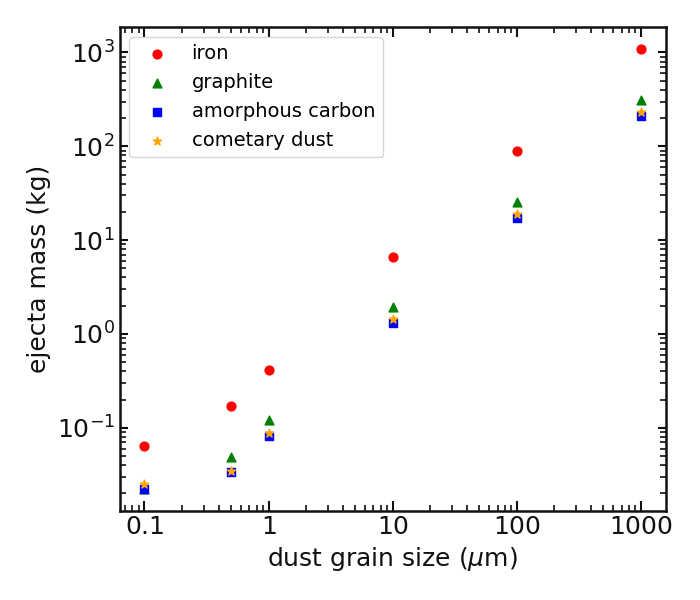}
    \caption{Variation of ejecta mass with dust grain size for strongly absorbing materials.}
    \label{mej_strong}
\end{figure}

\begin{figure}[H]
    \centering
    \includegraphics[width=0.7\textwidth]{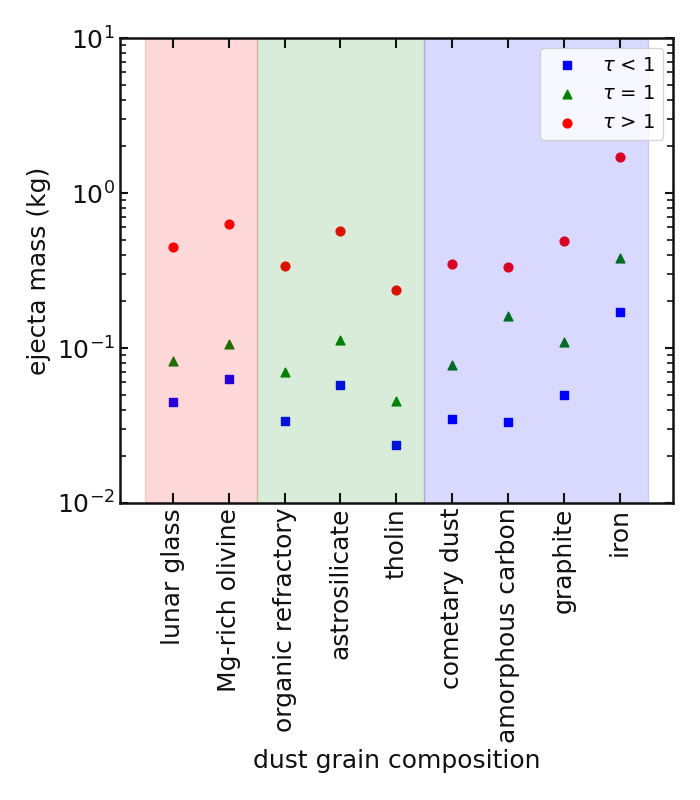}
    \caption{Variation of ejecta mass with the compositions at different optical depth.}
    \label{mej_allcomp_vartau}
\end{figure}

\begin{figure}[H]
    \centering
    \includegraphics[width=0.7\textwidth]{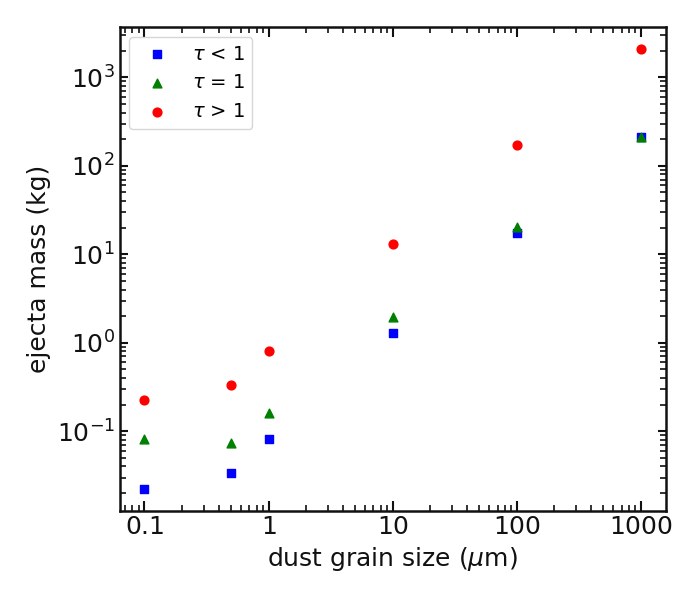}
    \caption{Variation of ejecta mass with size of amorphous carbon dust grains, at different optical depth.}
    \label{mej_amorcarbon_vartau}
\end{figure}

\begin{figure}[H]
    \centering
    \includegraphics[width=0.7\textwidth]{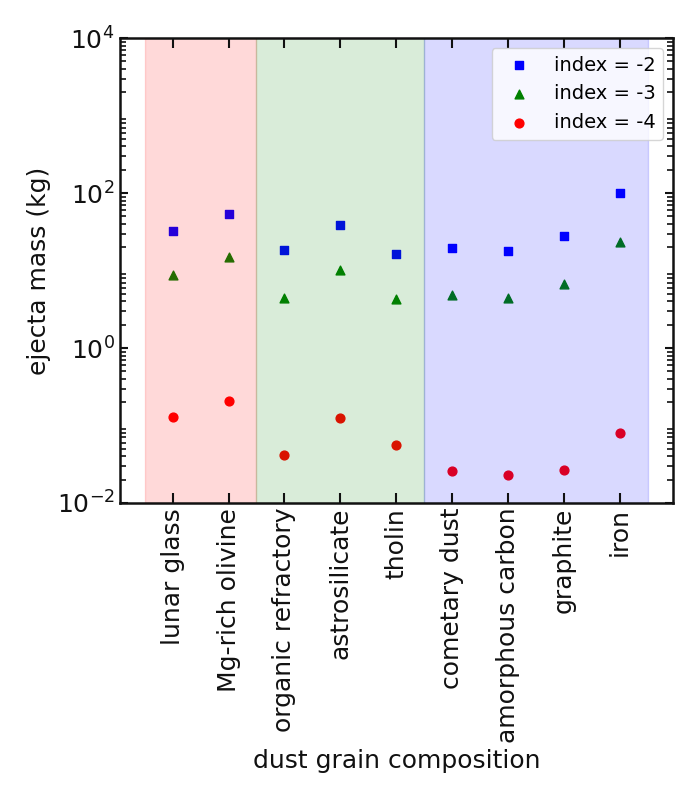}
    \caption{Variation of ejecta mass with the dust grain composition, considering grain-size distribution and constant number density across the ejecta curtain.}  \label{mej_allcomp_sizedistr_consttau}
\end{figure}

\begin{figure}[H]
    \centering
    \includegraphics[width=0.7\textwidth]{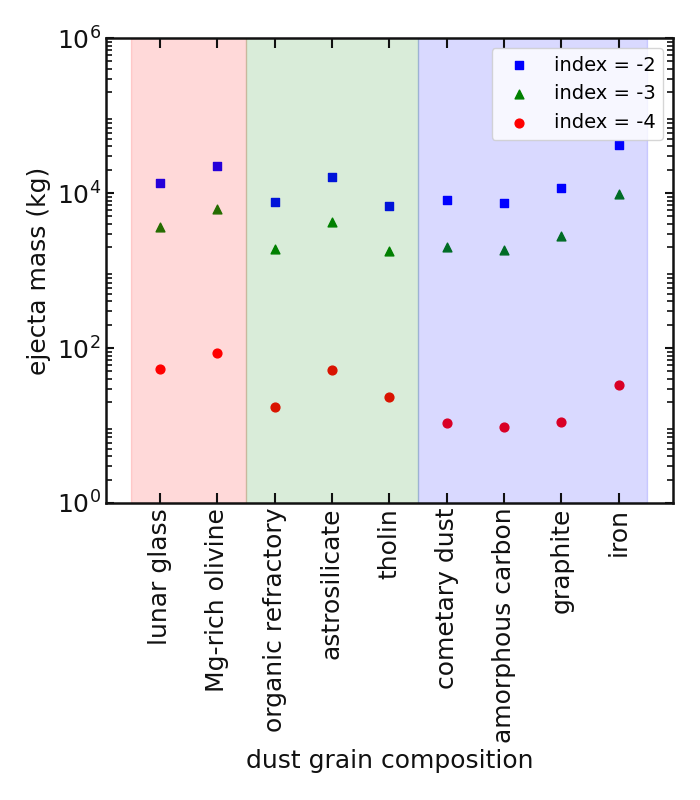}
    \caption{Variation of ejecta mass with the dust grain composition, considering grain-size distribution considering varying number density across the ejecta curtain.}
    \label{mej_allcomp_sizedistr_vartau}
\end{figure}

The total ejecta mass in the ejecta cone is calculated in the case of monodisperse and the size distribution of dust grains to understand influence of ejecta mass on the scattered intensity. The corresponding variation is shown in Figure \ref{mej_weak}, \ref{mej_moderate}, \ref{mej_strong}, \ref{mej_allcomp_vartau}, \ref{mej_amorcarbon_vartau}, \ref{mej_allcomp_sizedistr_consttau} and \ref{mej_allcomp_sizedistr_vartau}.


\clearpage
\bibliographystyle{sn-basic}
\bibliography{ecmodelbib}


\end{document}